\documentclass[sigconf]{acmart}
\AtBeginDocument{%
  }

\copyrightyear{2026}
\acmYear{2026}
\setcopyright{cc}
\setcctype{by}
\acmConference[WWW '26]{Proceedings of the ACM Web Conference 2026}{April 13--17, 2026}{Dubai, United Arab Emirates}
\acmBooktitle{Proceedings of the ACM Web Conference 2026 (WWW '26), April 13--17, 2026, Dubai, United Arab Emirates}
\acmPrice{}
\acmDOI{10.1145/3774904.3792724}
\acmISBN{979-8-4007-2307-0/2026/04}

\usepackage{amssymb}
\usepackage{algorithm2e}
\usepackage{multirow}
\usepackage{balance}
\usepackage{subfigure} 
\usepackage{hyperref}

\newcommand{\gpt}{ChatGPT Search\xspace}
\newcommand{\gemini}{Gemini-2.5 flash-grounding\xspace}
\newcommand{\pplx}{Perplexity-pro\xspace}
\newcommand{\claude}{Claude Sonnet 4 Search\xspace}

\newcommand{\proposed}{\textsc{AgenticShop}\xspace}
\begin{document}

\title{AgenticShop: Benchmarking Agentic Product Curation for~Personalized Web Shopping}

\author{Sunghwan Kim}
\authornote{Equal contribution}
\orcid{0009-0003-9514-812X}
\affiliation{%
  \department{Department of Artificial Intelligence}
  \institution{Yonsei University}
  \city{Seoul}
  \country{Republic of Korea}
}
\email{happysnail06@yonsei.ac.kr}

\author{Ryang Heo}
\authornotemark[1]
\orcid{0009-0008-1437-6435}
\affiliation{%
  \department{Department of Artificial Intelligence}
  \institution{Yonsei University}
  \city{Seoul}
  \country{Republic of Korea}
}
\email{ryang1119@yonsei.ac.kr}

\author{Yongsik Seo}
\orcid{0009-0006-3902-0553}
\affiliation{%
  \department{Department of Artificial Intelligence}
  \institution{Yonsei University}
  \city{Seoul}
  \country{Republic of Korea}
}
\affiliation{%
  \institution{ParamitaAI}
  \city{Seoul}
  \country{Republic of Korea}
}
\email{ysseo@yonsei.ac.kr}

\author{Jinyoung Yeo}
\orcid{0000-0003-3847-4917}
\affiliation{%
  \department{Department of Artificial Intelligence}
  \institution{Yonsei University}
  \city{Seoul}
  \country{Republic of Korea}
}
\email{jinyeo@yonsei.ac.kr}

\author{Dongha Lee}
\orcid{0000-0003-2173-3476}
\authornote{Corresponding author}
\affiliation{%
  \department{Department of Artificial Intelligence}
  \institution{Yonsei University}
  \city{Seoul}
  \country{Republic of Korea}
}
\affiliation{%
  \institution{ParamitaAI}
  \city{Seoul}
  \country{Republic of Korea}
}
\email{donalee@yonsei.ac.kr}

\renewcommand{\shortauthors}{Sunghwan Kim, Ryang Heo, Yongsik Seo, Jinyoung Yeo, \& Dongha Lee}

\begin{abstract}
The proliferation of e-commerce has made web shopping platforms key gateways for customers navigating the vast digital marketplace. 
Yet this rapid expansion has led to a noisy and fragmented information environment, increasing cognitive burden as shoppers explore and purchase products online. 
With promising potential to alleviate this challenge, agentic systems have garnered growing attention for automating user-side tasks in web shopping. 
Despite significant advancements, existing benchmarks fail to comprehensively evaluate how well agentic systems can curate products in open-web settings. 
Specifically, they have limited coverage of shopping scenarios, focusing only on simplified single-platform lookups rather than exploratory search. 
Moreover, they overlook personalization in evaluation, leaving unclear whether agents can adapt to diverse user preferences in realistic shopping contexts. 
To address this gap, we present \proposed, the first benchmark for evaluating agentic systems on personalized product curation in open-web environment. 
Crucially, our approach features realistic shopping scenarios, diverse user profiles, and a verifiable, checklist-driven personalization evaluation framework. 
Through extensive experiments, we demonstrate that current agentic systems remain largely insufficient, emphasizing the need for user-side systems that effectively curate tailored products across the modern web. 
\footnote{Code and dataset available at \url{https://github.com/happysnail06/AgenticShop}}
\end{abstract}





\maketitle

\section{Introduction}
\label{sec:introduction}
\begin{figure}[!t]
\centering
\includegraphics[width=0.97\linewidth]{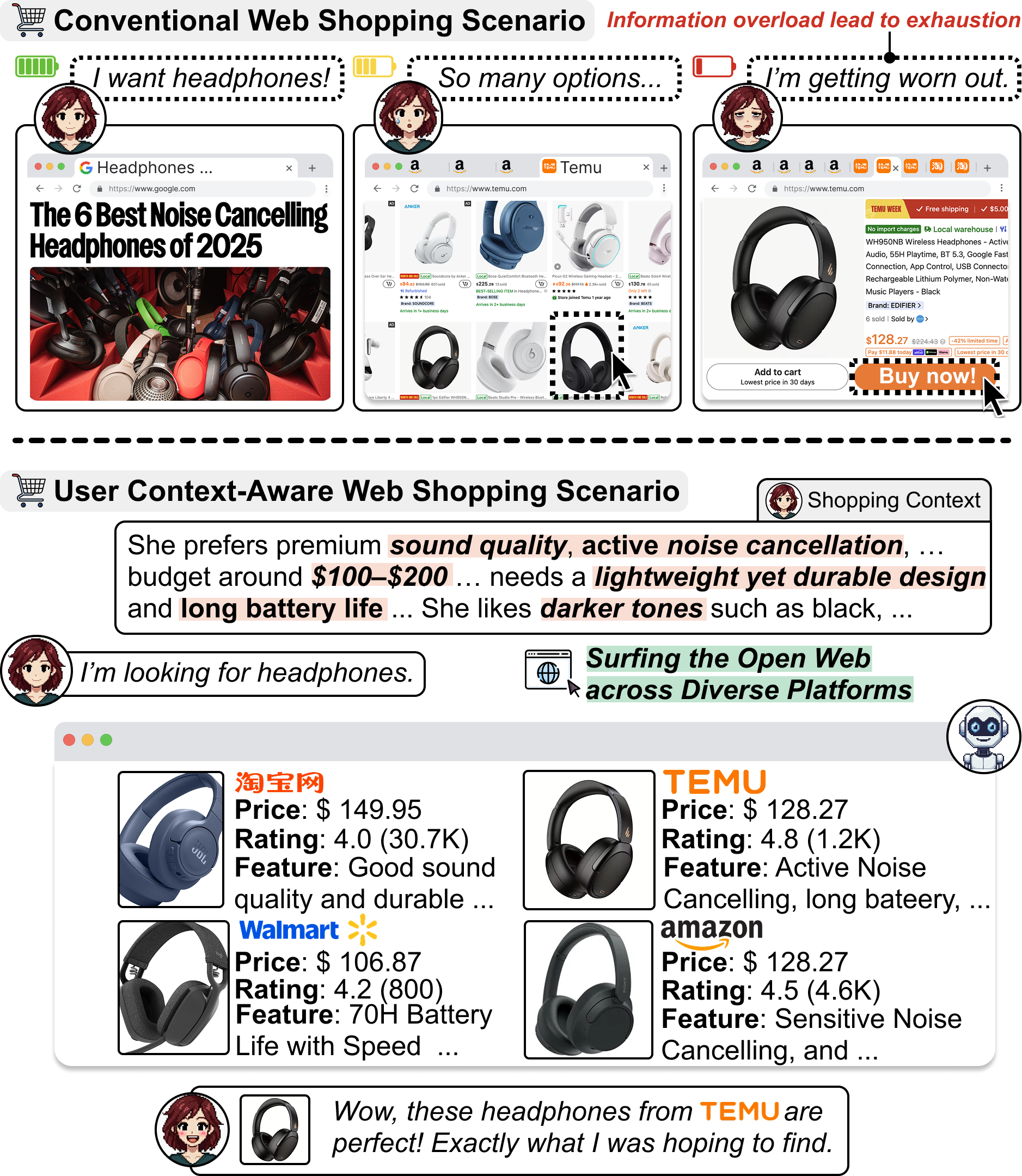}
\caption{
Conventional web shopping overwhelms users with fragmented information and fatigue (Upper), motivating a context-aware agent that curates personalized options from cross-platform evidence to ease the cognitive burden (Lower).
}
\label{fig:intro_fig}
\end{figure}

With the proliferation of e-commerce, web shopping platforms have become key gateways for customers navigating the vast digital marketplace. 
While offering unprecedented convenience, this rapid expansion has also produced a noisy and inconsistent information environment, where excessive promotional content, redundant descriptions, and biased reviews make it difficult for users to identify products that meet their needs~\cite{peng2021does, lv2022impact, zhang2023examining, heo2025can}.
This challenge is especially pronounced for high-involvement purchases, where users consult external sources such as blogs and communities, compare alternatives across multiple platforms, and make final purchase decisions based on price and offerings~\cite{darley2010toward, athukorala2016exploratory, kim2025user, dammu2025shopping} (Figure~\ref{fig:intro_fig}, Upper).

Despite this complexity, existing approaches in recommender systems and product search~\cite{duan2013supporting, van2016learning, zhang2019neural, dai2023contrastive, xu2024optimizing} have predominantly focused on improving product discovery within individual platforms, overlooking the exploratory nature of web shopping.
Users continue to invest substantial time searching and synthesizing fragmented information scattered across the web, suffering considerable cognitive burden and decision fatigue.
Motivated by this, developing a system capable of performing web-scale product exploration has become a fundamental challenge in the field of e-commerce. 
This calls for a shift from platform-specific solutions toward user-side systems that can integrate dispersed product information across the open web and deliver personalized curations (Figure~\ref{fig:intro_fig}, Lower).

With recent advances in large language models (LLMs), agentic systems have emerged as a promising paradigm for web-based tasks, making agent-driven shopping across the open web increasingly feasible. 
Two main approaches are prominent: \textit{Search-augmented LLMs}, which couple LLMs (e.g., ChatGPT) with search engines or retrievers to access up-to-date information~\cite{openai2023gpt4, perplexity2024, comanici2025gemini}, and \textit{Autonomous web agents}~\cite{deng2023mind2web, anthropic2024claudeuse, zheng2024gptvision, abuelsaad2024agent, zhouwebarena, gou2025navigating, openai2025operator}, which directly operate on live websites with the ability to browse, synthesize evidence, and execute actions in a human-like manner. 
To evaluate these emerging systems, a series of benchmarks have been introduced, ranging from simulated environments~\cite{yao2022webshop, deng2023mind2web, chen2024chatshop} to sandboxed applications~\cite{zhouwebarena, lu2024weblinx} and more recent live web settings~\cite{gou2025navigating, gou2025mind2web, xue2025an}, as well as domain-specific testbeds in e-commerce~\cite{jin2023amazon, peeters2025webmall, lyu2025deepshop, zhang2025functionality}.

\definecolor{darkgreen}{HTML}{006400}
\begin{table*}[t]
\caption{A comparison of \proposed to existing e-commerce related benchmarks.}
\centering
\label{tab:benchmark_comparison}
\resizebox{\textwidth}{!}{%
\begin{tabular}{lccccccc}
\toprule
\multirow{2.5}{*}{\textbf{Benchmark}} & \multicolumn{2}{c}{\textbf{Interaction Environment}} & \multicolumn{3}{c}{\textbf{Shopping Intent}} & \multicolumn{2}{c}{\textbf{Evaluation}} \\
\cmidrule(lr){2-3} \cmidrule(lr){4-6} \cmidrule(lr){7-8}
 & \textbf{Core Task} & \textbf{Open Web} & \textbf{Target Finding} & \textbf{Alternative Selection} & \textbf{Open Exploration} & \textbf{Method} & \textbf{Personalization} \\
\midrule
WebShop \cite{yao2022webshop} & Product Search & \color{red}$\times$ & \color{darkgreen}$\checkmark$ & \color{red}$\times$ & \color{red}$\times$ & Average Reward & \color{red}$\times$ \\
ShoppingBench \cite{wang2025shoppingbench} & Purchase Optimization & \color{red}$\times$ & \color{darkgreen}$\checkmark$ & \color{red}$\times$ & \color{red}$\times$ & Relevance Score & \color{red}$\times$ \\
PersonalWab  \cite{cai2025large} & Function Selection & \color{red}$\times$ & \color{darkgreen}$\checkmark$ & \color{red}$\times$ & \color{red}$\times$ & Answer Match & \color{red}$\times$ \\
WebMall  \cite{peeters2025webmall} & Product Comparison & \color{red}$\times$ & \color{darkgreen}$\checkmark$ & \color{darkgreen}$\checkmark$ & \color{red}$\times$ & Answer Match & \color{red}$\times$ \\
DeepShop \cite{lyu2025deepshop} & Product Search  & \color{darkgreen}$\checkmark$ & \color{darkgreen}$\checkmark$ & \color{red}$\times$  & \color{red}$\times$ & LLM-as-a-Judge & \color{red}$\times$\\
AmazonBench \cite{zhang2025functionality} & Function Execution  & \color{darkgreen}$\checkmark$ & \color{darkgreen}$\checkmark$ & \color{red}$\times$ & \color{red}$\times$ & LLM-as-a-Judge & \color{red}$\times$ \\
\midrule
\textbf{\proposed} (Ours) & Product Curation & \color{darkgreen}$\checkmark$ & \color{darkgreen}$\checkmark$ & \color{darkgreen}$\checkmark$ & \color{darkgreen}$\checkmark$ & LLM-as-a-Judge & \color{darkgreen}$\checkmark$ \\
\bottomrule
\end{tabular}%
}
\end{table*}
Despite these advances, current benchmarks fall short in evaluating product curation in realistic web shopping environments.
Specifically, there are two major limitations: \textbf{(1) Limited scope and coverage of shopping scenarios.} 
Web shopping is inherently \textit{exploratory}; while some users begin with a specific product in mind, many navigate fluidly across platforms to compare alternatives and refine their choices.
However, existing benchmarks reduce evaluation to single-platform lookups of predetermined items, failing to test whether agents can effectively discover and curate meaningful options from diverse sources.
\textbf{(2) Lack of personalization in evaluation.} 
User preferences in web shopping are \textit{highly diverse}; for example, some users prioritize delivery speed, while others care more about reviews.
Therefore, evaluation should extend beyond simple product search accuracy to rigorously account for such personalized attributes.
Nevertheless, no existing benchmark explicitly evaluates \textit{personalization}, leaving open whether agents can satisfy fine-grained, contextual factors that drive real shopping decisions.

Motivated by these gaps, we introduce \textbf{\proposed}, a novel benchmark designed to evaluate agentic systems on personalized product curation in open web environments.
Crucially, \proposed is distinguished in three ways:
\textbf{(1) Realistic shopping intents.} 
Building on prior user studies~\cite{su2018user, sondhi2018taxonomy}, we formalize three representative user goals---Target Finding (TF), where users request a specific product; Alternative Selection (AS), where users seek meaningful comparisons across options; and Open Exploration (OE), where users casually browse to discover appealing items.
These intents collectively reflect the exploratory behaviors of web shopping and define the basis for constructing test queries.
\textbf{(2) User profiles with authentic personalization.} 
We construct \textit{user shopping profiles} derived from real purchase histories and review texts. 
Specifically, each profile contains three components: a narrative describing the user's general shopping preferences and constraints, a query tailored to one of the three shopping intents, and a checklist categorizing the user's requirements into evaluable attributes.
Together, these components provide a robust foundation for evaluating personalization.
\textbf{(3) Checklist-driven personalization evaluation.}
We employ \textit{LLM-as-a-judge} to verify whether curated outputs meet the user's requirements. To achieve this, we systematically extract product information from the linked sources provided by the agents, grounding the judge's decisions in verifiable evidence for each checklist item. 
This approach provides a fine-grained and scalable way to measure how effectively agents satisfy individual user preferences across diverse shopping scenarios.

Our comprehensive experiments on \proposed reveal that current agentic systems remain insufficient for personalized product curation.
We find that query specificity contributes little to personalization, as the open web lacks the structured taxonomies present in specialized shopping platforms.
Performance also varies significantly across product domains, particularly when subjective visual or aesthetic preferences are key decision factors. 
We further observe that agentic systems struggle to consider core preference dimensions such as price or reviews, largely due to limited capacity for handling dynamic content and difficulty discovering products with adequate user reviews.
Overall, we show that effective personalized product curation depends on exploratory information seeking across diverse sources and careful comparison of alternatives.
To summarize, our contributions are as follows:
\setlength{\leftmargini}{10pt}
\setlength{\itemsep}{3pt}
\begin{itemize}
    \item We propose {\proposed}, the first benchmark designed for evaluating agentic systems on personalized product curation in open web shopping environments.
    
    \item We present an evaluation framework to comprehensively assess personalization, grounded in realistic shopping intents, diverse user profiles, and verifiable checklist-driven evaluation.

    \item We extensively evaluate both search-augmented LLMs and autonomous web agents on \proposed, thoroughly analyzing the key challenges they face and discussing future directions for advancing personalized product curation on the open web.

\end{itemize}

\section{Related Work}
\label{sec:relatework}
\subsection{Web-based Agentic Systems}
Recent advancements in Large Language Models (LLMs) have increasingly focused on developing systems that integrate web access to expand their capabilities. 
Early efforts took shape as search-augmented LLMs, such as WebGPT~\cite{nakano2021webgpt}, WebGLM~\cite{liu2023webglm}, and Perplexity Search~\cite{perplexity2024}, aimed at improving question answering by retrieving and synthesizing evidence from the web. 
In parallel, another line of research~\cite{gur2023real, koh2024tree, furuta2023multimodal, gu2024your} has focused on building autonomous web agents capable of solving highly interactive tasks.
These agents have rapidly evolved from text-based approaches~\cite{gur2023real, abuelsaad2024agent} to multimodal paradigms, as demonstrated by studies such as~\cite{zheng2024gptvision, he2024webvoyager, song2024beyond}.
Despite significant progress, the application of these agentic systems to real-world environment remains underexplored. 
E-commerce represents a particularly compelling domain where users face significant cognitive burden in exploring vast product catalogs, comparing options across multiple platforms, and making informed purchasing decisions.
Motivated by these challenges, this study provides a comprehensive evaluation of agentic systems' potential to automate user-side product exploration, integrating dispersed product information across the open web and presenting personalized curations.

\subsection{Agent Benchmarks}
Numerous benchmarks ~\cite{mialon2023gaia, koh2024visualwebarena, lu2024weblinx,tian2025mmina} have been introduced to evaluate the capabilities of web agents, progressing from static website snapshots~\cite{deng2023mind2web} to reproducible sandbox environments~\cite{koh2024visualwebarena}, and most recently to real-world live websites ~\cite{pan2024webcanvas, xue2025an, gou2025mind2web}.
Building on these developments, e-commerce has become a prominent domain for benchmarking web agents~\cite{jin2023amazon, chen2024chatshop, jin2024shopping, chen2025chineseecomqa}.
As shown in Table~\ref{tab:benchmark_comparison}, early efforts primarily focused on tasks such as locating and purchasing products~\cite{yao2022webshop}, whereas recent studies have shifted toward capturing more realistic shopping scenarios. 
For instance, ShoppingBench~\cite{wang2025shoppingbench} focuses on purchase optimization such as maximizing discounts and qualifying for free shipping.
WebMall~\cite{peeters2025webmall} evaluates web agents on comparison-shopping across four simulated websites.
DeepShop~\cite{lyu2025deepshop} increases search query complexity by injecting attributes, filters, and sorting criteria.
Amazon-Bench~\cite{zhang2025functionality} broadens task coverage to include account management, wishlist operations, and other functional capabilities.
Nevertheless, existing benchmarks still fall short of capturing the broader exploration process of real-world web shopping, remaining confined to product discovery or navigation within predefined platforms.
To address this, we introduce a new benchmark that comprehensively evaluates agentic systems across diverse and realistic shopping intents.

\subsection{Evaluation of Agentic Systems}
As research increasingly shifts toward online environments, evaluation has emerged as a central challenge in developing web agents.
While some research conduct manual human evaluation for assessing agent performance~\cite{he2024webvoyager}, its limited scalability has driven the development of automated approaches. 
In particular, prior studies~\cite{pan2024webcanvas, gou2025mind2web} introduced checkpoint-based evaluation, combining rule-based functions with LLM-as-a-judge ~\cite{zheng2023judging} for automated assessment.
Yet, these methods still rely on human annotation to identify critical checkpoints, leaving the scalability challenge only partially resolved.
To mitigate this, recent work has advanced toward fully human-agnostic evaluation. 
For example, WebJudge~\cite{xue2025an} leverages LLMs to identify key checkpoints and corresponding screenshots, and DeepShop~\cite{lyu2025deepshop} prompts LLMs to assess whether the outcome matches specific query requirements.
Nevertheless, evaluating agentic systems in e-commerce presents unique challenges, as task outcomes are often shaped by user-specific preferences and constraints.
While PersonalWAB~\cite{cai2025large} introduces the notion of personalization into web agents, it does not extend this concept to the evaluation stage.
In this work, we address this gap by proposing checklist-driven personalization evaluation framework.

\section{AgenticShop}
\label{sec:benchmark}
\begin{figure*}[t]
    \centering
    \includegraphics[width=\textwidth]{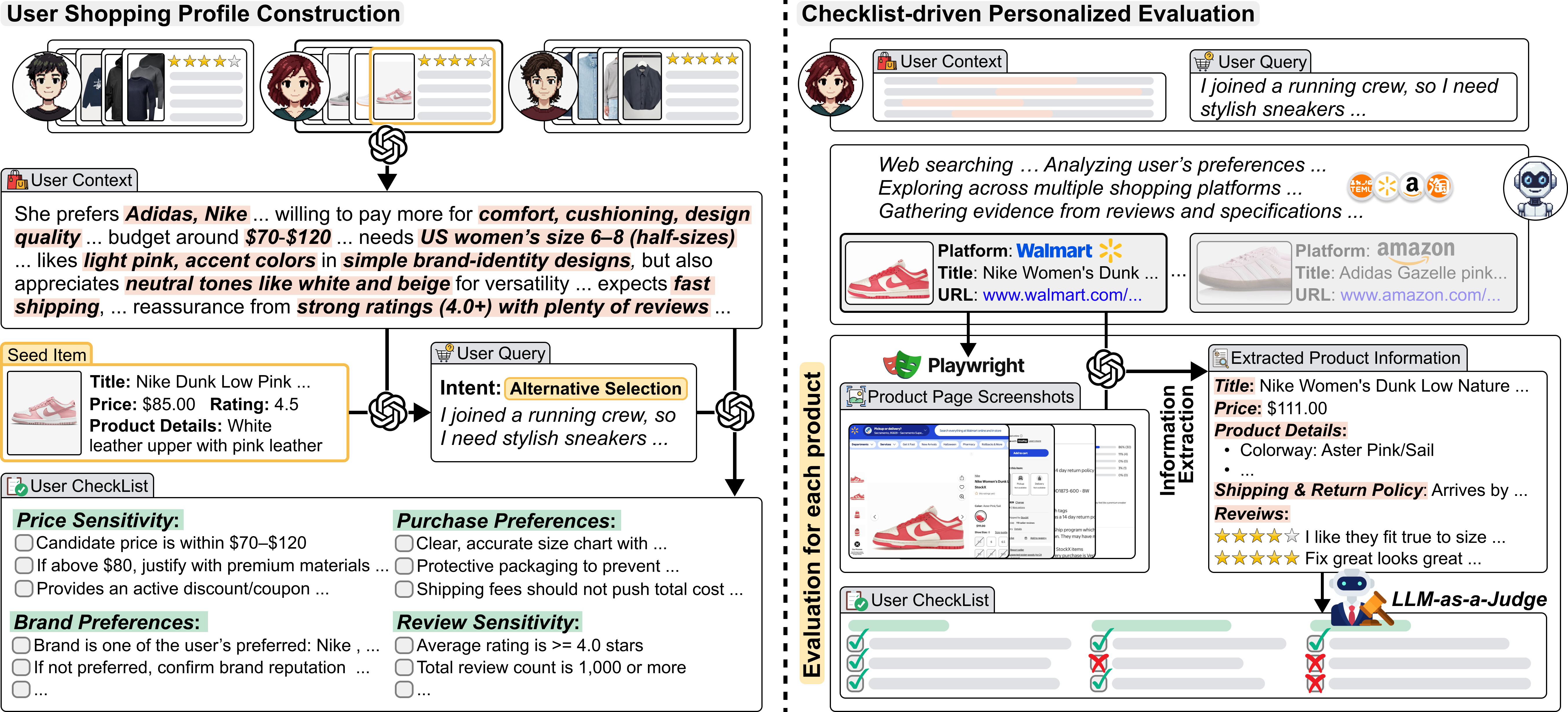}
    \caption{Overview of \proposed. For \textit{user profile construction}, real user purchase histories and review texts are used to build narrative-style user shopping contexts, from which intent-specific queries and personalized checklists are generated.
    For \textit{checklist-driven personalized evaluation}, LLM-as-a-judge verifies whether curated results satisfy each user’s shopping context, grounding its decisions in product information extracted from the linked pages of curated products provided by the agents.}
    \label{fig:main_fig}
\end{figure*}

In this section, we introduce \textbf{\proposed}, the first benchmark designed to comprehensively evaluate agentic systems on personalized
product curation in open-web settings (Figure~\ref{fig:main_fig}).
Our approach builds on three core principles:
\textbf{(1)} realistic scenarios derived from a taxonomy of user shopping intents, 
\textbf{(2)} diverse user profiles that reflect genuine human preferences, and 
\textbf{(3)} personalized evaluation methods demonstrating strong alignment with human judgments.

\subsection{Task Formulation}
Given a user $u \in \mathcal{U}$, the user context $X$ is represented as a narrative description of their personal shopping needs.
It encapsulates the user’s preferences, constraints, and decision-making tendencies inferred from their purchase history and reviews.
Based on this context, the user issues a query $q$ reflecting their current shopping intent.
The agentic system $\mathcal{A}$ then explores the open web $\mathcal{W}$ and generates a list of products:
\[
L = \mathcal{A}(q, X; \mathcal{W})
\]
where $L = \{r_1, r_2, \ldots, r_M\}$ denotes the {$M$ product candidates discovered for user $u$.}
The goal of the agentic system is to curate {products from $\mathcal{W}$} that best align with the user’s shopping context $X(u)$.

\subsection{User Shopping Intents \& Scenarios}
\label{sec:shopping_intents}
Following prior studies~\cite{su2018user, sondhi2018taxonomy}, we observe that shopping scenarios vary with user intent, reflecting distinct shopping behaviors and decision-making processes. 
Based on these insights, we group shopping intents into three types and evaluate agentic systems in corresponding scenarios.
We provide illustrations of shopping intents in Figure~\ref{fig:shopping_scenario_fig}, with descriptions of each type given below.
\setlength{\leftmargini}{10pt}
\setlength{\itemsep}{4pt}
\begin{itemize}
    \item \textbf{Target Finding (TF)}. 
    Users have a specific item in mind and seek with clear brand or product specifications.
    Queries may exhibit varying levels of specificity: (1) \textit{Explicit Title} queries that reference exact product names, (2) \textit{Attribute-Specific} queries that describe key product features, and (3) \textit{Brand-Categorical} queries that combine brand preferences with product type (e.g., ``\textit{ILLY capsule coffee}'').
    In this setting, the agent's task is to locate the product across different platforms and compare their offerings.
    
    \item \textbf{Alternative Selection (AS)}. 
    Users express a vague intent toward a certain product category without committing to a specific item.
    Users often issue a broad query such as ``\textit{Searching for earbud headphones}'' to start their exploration.
    In such cases, the level of user involvement becomes especially important, since different product domains demand varying degrees of decision-making effort.
    Specifically, \textit{Low-Involvement} products require moderate comparison of utility and aesthetic attributes.
    \textit{High-Involvement} products demand careful evaluation of technical specifications and feature trade-offs. \textit{Aesthetic-Driven} products depend heavily on subjective style preferences.
    The goal of the agent is to provide multiple alternatives that align with the user's preferences, allowing them to compare and make informed decisions.
    
    \item \textbf{Open Exploration (OE)}. 
    Unlike the above scenarios, users may not issue any explicit query in Open Exploration.
    Instead, they casually browse the open web to explore potentially appealing items.
    Here, the agentic system's role shifts to proactive curation, presenting diverse items across different domains (e.g., apparel, furniture, or snacks) that potentially satisfy the user's interests.
\end{itemize}
\subsection{Benchmark Construction}
\label{sec:benchmark_construction}
We use the Amazon Review ~\cite{hou2024bridging} as our source dataset, as it provides rich user interaction data spanning extensive product catalogs. 
By systematically sampling across domains and users, we create user profiles that are both diverse and comprehensive for our benchmark.

\subsubsection{Domain Selection.}
We choose domains where each intent type's characteristics are most evident \cite{sondhi2018taxonomy}.
Specifically, Target Finding is represented by \textit{Grocery \& Gourmet Food}, where shoppers typically search for specific products already in mind. 
Alternative Selection is modeled using \textit{Home \& Kitchen, Electronics, and Clothing, Shoes \& Jewelry}, corresponding to \textit{Low-Involvement, High-Involvement, and Aesthetic-Driven} shopping scenarios, respectively. 
Finally, Open Exploration aggregates all domains to assess the agent's ability to adapt curation strategies across diverse domains.

\subsubsection{User Sampling.}
In our framework, each user profile serves as the foundation for a distinct evaluation scenario, representing a realistic shopping task within a specific product category. 
To ensure user profiles reflect consistent and comprehensive preferences, we leverage the Amazon Review product taxonomy, where each purchased item in the user interaction history is tagged with specific subcategories within the domain.
Specifically, we apply a three-step filtering process.
First, we select users with at least 100 reviews to ensure sufficient interaction history. 
Second, we identify each user’s core shopping interest by analyzing their most frequently purchased product subcategories.
We refer to this most representative subcategory as the user’s \textit{target product type}.
Finally, we retain only purchase histories related to their target product type, creating focused profiles that reflect genuine, category-specific preferences.
For example, if a user's history is from Electronics domain and concentrates in a product type such as headphones, we initiate an {evaluation scenario} where the user is shopping for a headphone.
 
\subsubsection{User Context}
\label{sec:user_context}
After selecting users with purchase histories focused on their target product type, {we ask an LLM to generate a descriptive narrative that encapsulates the key factors contributing to each user’s decision-making process ~\cite{kim2024stop, kim2025towards}.} 
We leverage GPT-5-mini to achieve this, with both product metadata 
and associated reviews provided as input.
These narratives are then used as references, allowing the agent to explore and curate products that align with the user context. 
In doing so, we aim to evaluate the ability of agents to achieve personalization in the open web environment.

\subsubsection{User Query}
\label{sec:user_query}
Following the taxonomy of user shopping intents described in Section~\ref{sec:shopping_intents}, we generate queries tailored to each scenario.
We randomly select a \textit{seed item} from the user's filtered purchase history to create a situation where the user seeks a specific product.
To reflect the varying levels of query specificity observed in Target Finding scenarios, we generate three query variations: Explicit Title, Attribute-Specific, Brand-Categorical.
This variation allows us to assess how agents handle different degrees of query specificity when searching target products.
For Alternative Selection scenarios, users have only a product type in mind.
Thus, we generate queries expressing broad intent toward the target product type.
Importantly, these queries omit specific attributes like brands that appear in Target Finding queries.
Lastly, for Open Exploration scenarios, no explicit query is provided.
Instead, we task the agent with proactively curating items the user may be interested in.

\subsubsection{{User Checklist}}
Effective personalization evaluation should be multifaceted, atomic, and verifiable.
We capture this by designing a checklist spanning six dimensions.: \textit{brand preferences} (inclination toward or disinterest in specific brands); \textit{price sensitivity} (responsiveness to price variations and budget constraints); \textit{review sensitivity} (degree of dependence on product ratings and user reviews); \textit{functional requirements} (essential technical specifications or features); \textit{aesthetic preferences} (preferences for visual styles, design elements, or product appearance); and \textit{purchase preferences} (expectations regarding sellers, shipping, and services).
We prompt GPT-5-mini to generate this checklist, providing a comprehensive and fine-grained basis for evaluating personalization quality.


\subsection{Personalized Evaluation}
Evaluating agentic systems in online environments poses inherent challenges. 
Particularly in e-commerce, this process centers on two key aspects: (1) ensuring verifiable and objective evaluation, and (2) properly accounting for personalization across diverse users.

\subsubsection{Information Extraction}
Given a user’s context and intent-specific query, the agentic system $\mathcal{A}$ explores multiple shopping platforms to curate products that align with the user’s preferences and constraints.
To ensure reliable and faithful evaluation, we extract all available product information from the product pages of the agent-curated items.
Specifically, for each curated product $r_i$, we collect all presented information such as brand, price, descriptions, delivery options, and user reviews.
To capture dynamically loaded content that may not appear in the initial page and to suppress noise from promotional elements (e.g., advertisements and cross-sell banners), 
we employ the \textit{Playwright} library\footnote{\url{https://playwright.dev/python/docs/intro}} to render the entire webpage and obtain a set of screenshots $\mathcal{S} = \{ S_1, S_2, \dots, S_n \}$, each corresponding to a fully rendered viewport segment of the product page for $r_i$.
We then provide both the product’s URL, denoted by $\text{URL}(r_i)$, and the screenshot set $\mathcal{S}{_i}$ as joint inputs to an LLM-based extractor $\mathcal{M}_{\text{extract}}$ equipped with web-searching capabilities, which performs structured information extraction as follows:
\[
\mathcal{I}(r_i) = \mathcal{M}_{\text{extract}}\big(\text{URL}(r_i), \mathcal{S}\big),
\]
Here, $\mathcal{I}(r_i)$ denotes all the product information extracted.
This multimodal approach enables robust extraction across e-commerce platforms with heterogeneous layouts and content structures, providing reliable objective evidence for personalized evaluation.

\subsubsection{Evaluation}
Finally, we leverage an \textit{LLM-as-a-judge} framework to evaluate curated products against the personalized checklist, based on the extracted product information.
Specfically, we prompt the LLM (i.e., GPT-5-mini) to perform binary verification for each checklist criterion, explicitly determining whether a product satisfies or violates it.
Formally, let $u$ be associated with personalized checklist $\mathcal{C} = \{C_1, C_2, \ldots, C_K\}$ consisting of $K$ criteria, and let $\mathcal{I}(r_m)$ denote the metadata extracted for a specific product $r_m$ from a list of M products.
For each product–criterion pair $(\mathcal{I}(r_m), C_k)$, the \textit{LLM-as-a-judge} outputs a binary decision indicating whether the product satisfies the given criterion (assigned a value of 1) or fails to meet it (assigned a value of 0).
We then compute the average proportion of satisfied criteria across all users, products, and checklists.
We define the overall curation score as:
\[
Score= \frac{1}{|\mathcal{U}| M K} 
\sum_{u \in \mathcal{U}} \sum_{m=1}^{M} \sum_{k=1}^{K} J(\mathcal{I}(r_m), C_k),
\]
where $J(r_m, C_k)$ denotes the \textit{LLM-as-a-judge} decision.
This integrated evaluation provides a fine-grained and scalable way to measure how effectively agentic systems satisfy individual user preferences and requirements across diverse web shopping scenarios.

\section{Experiments}
\label{sec:experiments}
\medskip
\noindent\textbf{Baselines.}
We conduct extensive experiments on two types of agentic systems:
\textbf{(1) Search-augmented LLMs}, including \gpt~\cite{openai2024chatgptsearch}, \claude~\cite{anthropic2024websearch}, \gemini~\cite{comanici2025gemini}, and \pplx~\cite{perplexity2024}, that generate responses by incorporating information retrieved through web search tools; and
\textbf{(2) Autonomous web agents}, including Agent-E~\cite{abuelsaad2024agent}, SeeAct~\cite{zheng2024gptvision}, Web Voyager~\cite{he2024webvoyager}, and Browser Use~\cite{muller2024browseruse}, that directly interact with live web interfaces  through browsing and navigation.

\subsection{Meta-Evaluation}
To verify the robustness of our evaluation, we perform meta evaluation focusing on two aspects: (1) the reliability of the checklist, and (2) the alignment between the LLM-as-a-judge and human judges.

\subsubsection{{Checklist Reliability Analysis}}
We first examine whether the generated checklists faithfully capture user preferences and can thus serve as reliable evaluation criteria for personalized product curation.
To this end, we conduct a preference alignment experiment using user ratings as ground truth, as they provide a direct and quantifiable signal of preference intensity.
For each evaluation instance, an LLM (i.e., GPT-5-mini) is provided with three inputs: a user checklist, one low-rated item, and one high-rated item drawn from the same user’s interaction history.
The model is then instructed to determine which product better reflects the user’s preferences.
As baselines, we also test alternative user representations—such as queries, purchase histories, and user context—in place of the checklist.
As reported in Table~\ref{tab:user_alignment_table_transposed}, the checklist achieves the highest alignment accuracy (0.81 on average), consistently outperforming all other baselines across domains.
This strong alignment with real user ratings validates the checklist as a reliable criterion for evaluating how effectively agentic systems address personalization.
\begin{table}[t]
\centering
\small
\caption{Alignment analysis of how effectively the checklist and alternative user representations capture user preferences across domains. The evaluation is performed on 250 samples.}
\label{tab:user_alignment_table_transposed}
\resizebox{0.85\columnwidth}{!}{
\begin{tabular}{lcccccc}
\toprule
\multirow{1}{*}{\textbf{Method}} & \textbf{Food} & \textbf{Home} & \textbf{Elec.} & \textbf{Fashion} & \textbf{All} & \textbf{Avg.} \\
\midrule
Query & 0.52 & 0.52 & 0.56 & 0.50 & 0.60 & 0.54 \\
Histories & 0.68 & 0.68 & 0.72 & 0.74 & 0.66 & 0.70 \\
Context & \underline{0.70} & \underline{0.74} & \underline{0.80} & \underline{0.78} & \underline{0.92} & \underline{0.78} \\
\midrule
Checklist & \textbf{0.72} & \textbf{0.76} & \textbf{0.82} & \textbf{0.80} & \textbf{0.94} & \textbf{0.81} \\
\bottomrule
\end{tabular}
}
\end{table}

\subsubsection{LLM-as-a-judge Reliability Analysis}
To further verify the robustness of our evaluation protocol, we conduct a human study comparing the LLM’s evaluations with those of human judges.
Specifically, we sample 50 instances from the evaluation set, each containing a product URL, extracted product information, and its corresponding checklist.
Human evaluators are instructed to review each product by opening the URL, examining the product information, and assessing the checklist accordingly.
Each instance is independently annotated by two human evaluators following the same evaluation guidelines used for the LLM-as-a-judge.
For comparison, we include two baseline settings: (1) w/o product info, where the LLM judges using only the URL and screenshots, (2) w/o screenshots, where the LLM judges solely based on the product URL.
As shown in Table~\ref{tab:human_eval}, our approach achieves the highest correlation with human assessments, demonstrating superior reliability and alignment over baseline methods.
In contrast, removing extracted information significantly reduces agreement, with the sharpest drop observed when visual evidence is also excluded.
This finding highlights the critical role of accurate information extraction in providing verifiable evidence, enabling the model to reference factual content and perform reliable evaluations.
We also provide detailed results for the six checklist dimensions in Appendix ~\ref{appendix:checklist_meta_eval}.

\begin{table}[t]
\centering
\scriptsize
\caption{Correlation between LLM-as-a-judge and human judgments across evaluation approaches. \textit{(p-value < 0.05)}}
\resizebox{0.89\columnwidth}{!}{
\begin{tabular}{lccc}
\toprule
\textbf{Criterion} & \textbf{Cohen's $\kappa$}  & \textbf{Spearman's $\rho$} \\
\midrule
\textbf{Ours}      &  \textbf{0.7320} &  \textbf{0.7425} \\
w/o product info. & 0.4395 & 0.4405 \\
w/o screenshots (URL Only) & 0.3792 & 0.3885 \\
\midrule
Inter-Human Correlation           & 0.7736 & 0.7736 \\
\bottomrule
\end{tabular}       
}
\label{tab:human_eval}
\end{table}

\section{Results and Discussion}
In this section, we present the main findings of our study, focusing on how agentic systems perform in personalized product curation.

\setlength{\tabcolsep}{2pt}
\renewcommand{\arraystretch}{1.1}
\newcommand{\GradDelta}[1]{#1}
\begin{table*}[ht]
\caption{Personalized curation performance comparison of various agentic systems across shopping intents and their queries.}
\centering
\scriptsize
\resizebox{0.88\linewidth}{!}{
\begin{tabular}{p{2.5cm}ccccccc}
\toprule
\multirow{2.5}{*}{\textbf{Agent}}
& \multicolumn{3}{c}{\textbf{TF}} 
& \multicolumn{3}{c}{\textbf{AS}} 
& \multicolumn{1}{c}{\textbf{OE}} \\
\cmidrule(lr){2-4} \cmidrule(lr){5-7} \cmidrule(lr){8-8}
& \textbf{Exp. Title}
& \textbf{Attr. Spec.}
& \textbf{Brand Cat.}
& \textbf{High Inv.}
& \textbf{Low Inv.} 
& \textbf{Aesthe. Driv.} 
& \textbf{Cas. Brow.} \\
\midrule
\multicolumn{8}{l}{\textbf{\textit{Search-augmented LLMs}}} \\
\gpt         &  \underline{37.81} &  \underline{35.35} &  \underline{34.79} & \textbf{30.13} &  \textbf{29.47} &  \textbf{27.55} &  {14.51} \\
\claude      &  \textbf{37.93} &  \textbf{36.44} &  \textbf{34.96} &  \underline{29.66} &  \underline{29.08} &  \underline{20.58} &  \textbf{21.20} \\
\gemini      &  22.29 &  21.52 &  23.60 &  18.04 &  19.72 &  16.62 &  \underline{16.06} \\
\pplx        &  22.85 &  22.04 &  21.03 &  19.95 &  22.57 &  14.36 &  14.36 \\
\midrule
\multicolumn{8}{l}{\textbf{\textit{{Autonomous Web Agents}}}} \\
Agent-E      &  19.75 &  20.65 &  14.85 &  16.15 &  21.76 &  \underline{19.98} &  13.56 \\
SeeAct       &  \textbf{37.13} &  \textbf{36.32} &  \textbf{34.92} &  \underline{25.50} &  \textbf{27.47} &  \textbf{20.80} &  \textbf{20.79} \\
Web Voyager  &  \underline{34.25} &  30.34 &  27.22 &  \textbf{25.87} &  \underline{25.68} &  18.76 &  15.57 \\
Browser Use  &  31.48 &  \underline{34.52} &  \underline{30.58} &  22.61 &  23.45 &  16.24 &  \underline{18.41} \\
\bottomrule
\end{tabular}
}
\label{tab:main_tasks}
\end{table*}
\subsection{RQ1: How effectively can agentic systems curate personalized products?}
\medskip
\noindent\textbf{Agentic systems struggle with personalized product curation.}
As shown in Table~\ref{tab:main_tasks}, we observe that the performance of agentic systems remain modest across all scenarios, indicating substantial room for improvement in personalized product curation.
Among search-augmented LLMs, \gpt and \claude exhibit the strongest performance, yet achieve only around 30-35\% across scenarios. 
In contrast, \gemini and \pplx show considerably weaker results, likely due to differences in search integration quality and underlying reasoning capabilities~\cite{miroyan2025search, kim2025bespoke}. 
For autonomous web agents, Agent-E shows considerably weaker capabilities compared to other baselines.
This disparity can be attributed to Agent-E's architectural emphasis on DOM distillation over multimodal understanding, where lengthy textual contexts potentially lead to higher error rates in completing the task  (Table~\ref{tab:combined_error_page_analysis}).
Lastly, all models show significantly lower performance in Open Exploration.
This difficulty stems from the need to balance relevance with novelty across diverse domains, and the lack of explicit user queries to guide their product search strategy.

\medskip
\noindent\textbf{Limited impact of query specificity.}
As shown in Table~\ref{tab:main_tasks}, we observe marginal performance differences across the three query variations in Target Finding scenarios, with only slight improvements when queries include more detailed specifications. 
In search-augmented LLMs, the models inherently engage in repeated query rewriting to explore information from different perspectives and integrate the retrieved results.
However, these reformulations often produce overly complex or inconsistent queries, making it difficult to obtain complementary search results and synthesize information effectively. 
Autonomous web agents also show a similar pattern by issuing a single long structured query at the start, which may lead to suboptimal search results. 
This suggests that enhancing search queries alone may not substantially improve personalized curation, as the open web often lacks the standardized taxonomies and structured catalogs found in commercial shopping platforms. 

\begin{figure}[t]
    \centering
    \includegraphics[width=0.95\linewidth]{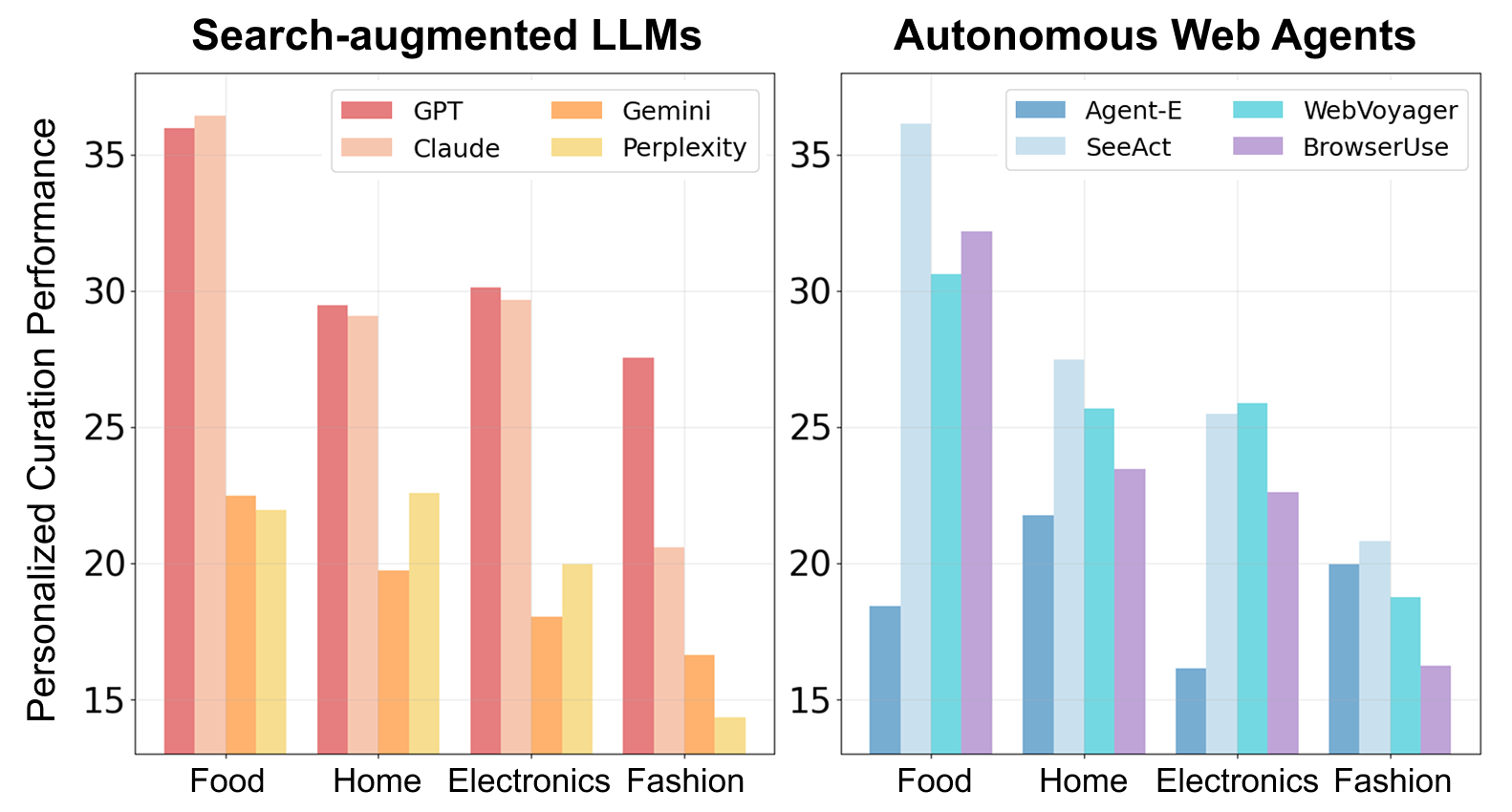}
    \caption{Performance across product domains.}
    \label{fig:domain_analysis} 
\end{figure}

\setlength{\tabcolsep}{2pt}
\begin{table}[t]
\caption{Comparison of Hit Ratios across query specificity.}
\centering
\scriptsize
\resizebox{0.93\linewidth}{!}{
\begin{tabular}{p{2.5cm}ccc}
\toprule
\multirow{1}{*}{\textbf{Agent}}
& \textbf{Exp. Title} & \textbf{Attr. Spec.} & \textbf{Brand Cat.} \\
\midrule
\multicolumn{4}{l}{\textbf{\textit{Search-augmented LLMs}}} \\
\gpt        &  0.70 &  0.44 &  0.60 \\
\claude        &  \textbf{0.86} &  \underline{0.50} &  \underline{0.60} \\
\gemini       &  \underline{0.72} &  \textbf{0.54} &  \textbf{0.66} \\
\pplx  &  0.46 &  0.38 &  0.48 \\
\midrule
\multicolumn{4}{l}{\textbf{\textit{{Autonomous Web Agents}}}} \\
Agent-E                &  0.24 &  0.10 &  0.14 \\
SeeAct                 &  \textbf{0.88} &  \textbf{0.48} &  \underline{0.28} \\
Web Voyager            &  0.70 &  0.36 &  \textbf{0.32} \\
Browser Use            &  \underline{0.76} &  \underline{0.36} &  0.24 \\
\bottomrule
\end{tabular}
}
\label{tab:tf_only}
\end{table}
\medskip
\noindent\textbf{Performance varies across product domains.}
As illustrated in Figure~\ref{fig:domain_analysis}, we observe that Electronics (i.e., \textit{High-Involvement})  generally pose greater challenges than Home \& Kitchen (i.e., \textit{Low-Involvement}), with most baselines performing better on the latter scenarios.
However, this gap narrows for more capable models. 
\gpt and \claude show comparable performance across both domains, suggesting that stronger model capabilities enable agentic systems to better handle complex product curations that demand domain expertise.
In contrast, for both search-augmented LLMs and web agents, performance drops significantly in Fashion (i.e., \textit{Aesthetic-Driven}), with the sharpest decline observed in \claude.
This reveals a critical limitation in current systems' ability to personalize curation when subjective visual or aesthetic preferences are the primary decision factors.

\subsection{RQ2: What makes it challenging for agentic systems to perform personalized curation?}
\label{sec:rq2}

\medskip
\noindent\textbf{Search capability does not translate to personalization quality.}
To examine whether low personalization performance stems from agents failing to locate relevant products, we conduct experiments using the three query variations.
Specifically, we evaluate product search capability independently from personalization quality, considering a product successfully located if it matches the target's brand and type, regardless of user checklist alignment.
As shown in Table~\ref{tab:tf_only}, most agents demonstrate reasonable search performance, with \claude achieving 86\% accuracy with \textit{Explicit Title} queries.
This shows that accurate product search does not necessarily translate into effective personalization.
For search-augmented LLMs, the limitation may stem from their reliance on static retrieval results, which limits their ability to progressively improve personalization.
In contrast, for autonomous web agents, the bottleneck appears to arise from their inability to recover after landing on a suboptimal product page—they tend to terminate early without engaging in further exploration.

\medskip
\noindent\textbf{Attribution errors persist across all evaluated systems.}
We analyze attribution errors that lead to task failures across all agentic systems. 
As shown in Table~\ref{tab:combined_error_page_analysis}, even off-the-shelf models like ChatGPT Search fail to provide correct product links, hallucinating approximately 20\% of product references. 
We identify two prevalent error patterns: (1) hallucinated pages where models generate invalid URLs (e.g., non-existent ASINs on Amazon), resulting in 404 errors, and (2) irrelevant links where models return generic search result pages, product listings, or other non-product pages.
Notably, invalid pages dominate the error cases, indicating that current systems lack robust grounding mechanisms for verifiable product attribution. 
Interestingly, when excluding these error cases and recalculating personalization scores on successful runs only, autonomous web agents outperform search-augmented LLMs (Table ~\ref{tab:main_tasks_full}). 
\begin{table}[t]
\caption{Comprehensive analysis of agentic systems' curated products, combining error and page-level statistics.}
\centering
\scriptsize
\resizebox{\columnwidth}{!}{
\begin{tabular}{lccc|cc}
\hline
\textbf{Agent}
& \begin{tabular}[c]{@{}c@{}}\textbf{Failure}\\\textbf{Rate} $\downarrow$ (\%)\end{tabular}
& \begin{tabular}[c]{@{}c@{}}\textbf{Irrelevant}\\\textbf{Page} (\%)\end{tabular}
& \begin{tabular}[c]{@{}c@{}}\textbf{Invalid}\\\textbf{Page} (\%)\end{tabular}
& \begin{tabular}[c]{@{}c@{}}\textbf{No Review}\\\textbf{Page} $\downarrow$ (\%)\end{tabular}
& \begin{tabular}[c]{@{}c@{}}\textbf{Out of}\\\textbf{Budget} $\downarrow$ (\%)\end{tabular} \\
\hline
\multicolumn{6}{l}{\textbf{\textit{Search-augmented LLMs}}} \\
ChatGPT Search & \textbf{24.76} & 28.35 & 71.65 & \textbf{0.35} & \textbf{0.70} \\
Claude Sonnet 4 Search & \underline{30.00} & 35.56 & 64.44 & \underline{0.40} & \underline{0.70} \\
Gemini-2.5 flash-grounding & 51.14 & 87.64 & 92.36 & 0.53 & 0.72 \\
Perplexity-pro & 56.10 & 12.39 & 87.61 & 0.52 & 0.74 \\
\hline
\multicolumn{6}{l}{\textbf{\textit{Autonomous Web Agents}}} \\
Agent-E & 59.10 & 28.39 & 71.61 & \textbf{0.16} & \textbf{0.65} \\
SeeAct & \textbf{32.66} & 17.84 & 82.16 & 0.36 & 0.71 \\
Web Voyager & 41.47 & 24.37 & 75.63 & \underline{0.31} & \underline{0.67} \\
Browser Use & \underline{39.14} & 13.14 & 86.86 & 0.33 & 0.73 \\
\hline
\end{tabular}}
\label{tab:combined_error_page_analysis}
\end{table}

\begin{figure}[t] 
    \centering
    \includegraphics[width=0.99\linewidth]{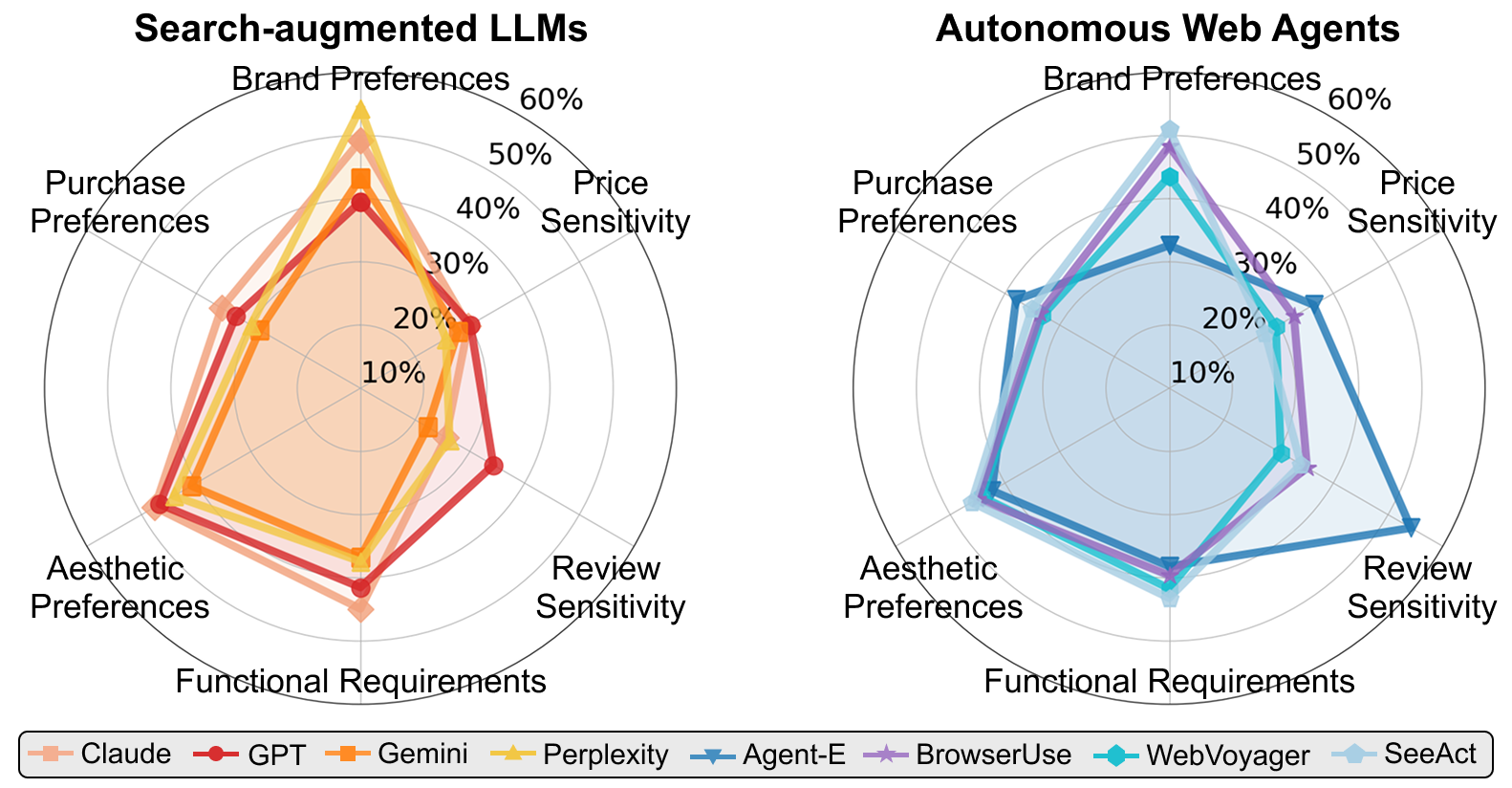}
    \caption{Performance across the six checklist dimensions.}
    \label{fig:checklist_dimension_analysis}
\end{figure}

\medskip
\noindent\textbf{{Agentic systems struggle with dynamic contents and reviews.}}
To identify where personalization succeeds and fails, we analyze performance across individual checklist dimensions.
As shown in Figure~\ref{fig:checklist_dimension_analysis}, while systems effectively address brand preferences, they exhibit significant limitations in capturing review sensitivity and price sensitivity.
To investigate the underlying causes, we examine product attributions alongside the evaluated checklist.
The results are reported in Table~\ref{tab:combined_error_page_analysis}.
For review sensitivity, we find that most systems curate products that lack customer reviews (indicated by the ``No Review Page'' field).
This suggests that the core limitation lies not in interpreting reviews, but in locating review-rich products during exploration.
For price sensitivity, we analyze the evaluated checklists and find that approximately 70\% of failures stem from out-of-budget recommendations.
We attribute this to two possible factors.
First, web prices are dynamic and fluctuate frequently across product variants and purchase options. 
Second, users often consider multiple cost components when setting budgets (e.g., "under \$50 including shipping"), but current agentic systems struggle to aggregate scattered pricing information such as base prices, shipping costs, taxes or discounts to calculate accurate final costs.

\subsection{RQ3: How can we enhance the personalized curation capabilities of agentic systems?}
\label{sec:insight}
\begin{figure}[t]
    \centering
    \includegraphics[width=0.88\linewidth]{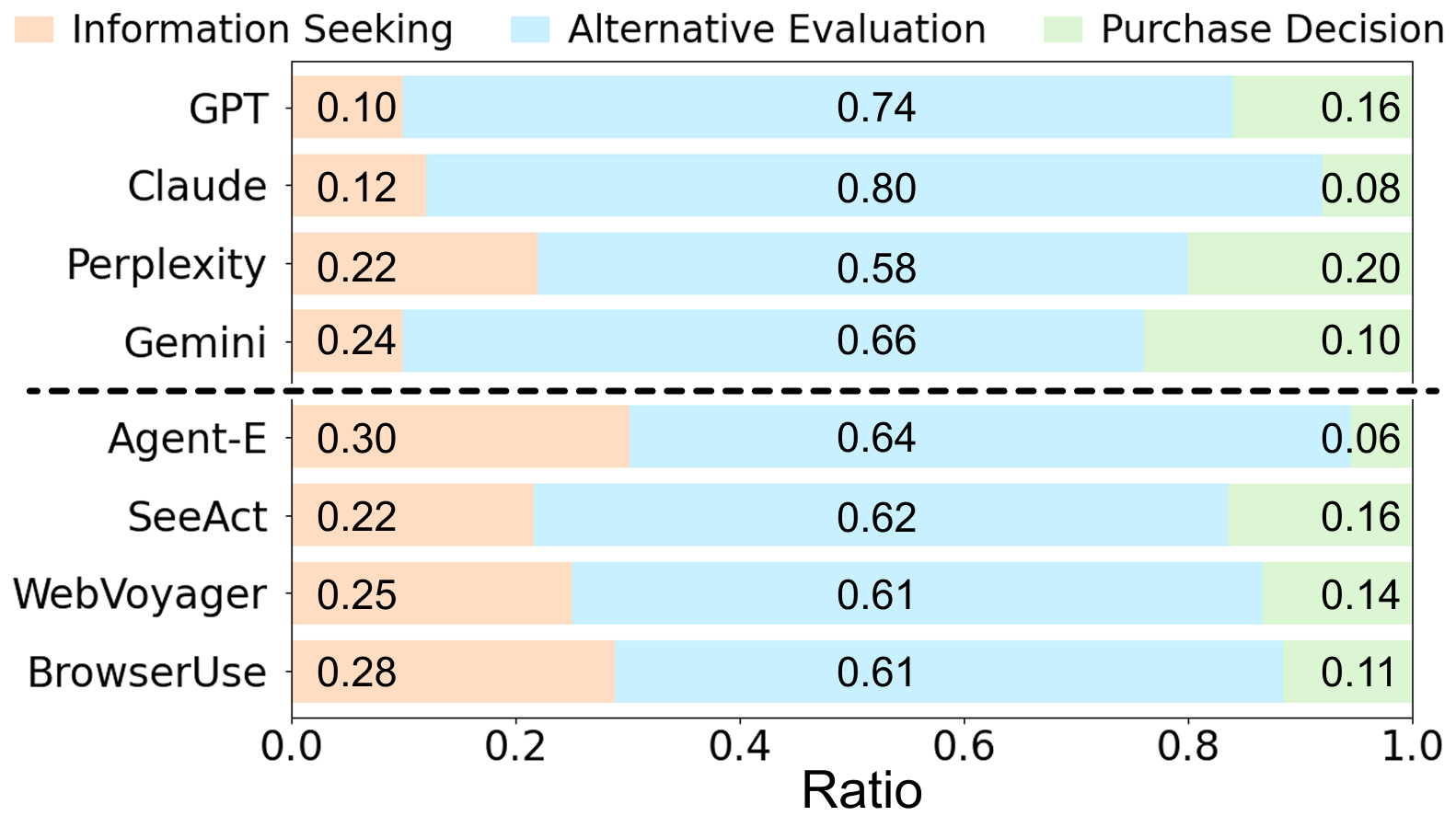}
    \caption{Intent distribution of agentic systems.}
    \label{fig:action_analysis}
\end{figure}

\medskip
\noindent\textbf{Alternative evaluation strongly correlates with personalization performance.}
To gain deeper insights into what drives personalized curation performance, we examine how systems behave during the curation process. 
Inspired by consumer decision-making models~\cite{panwar2019consumer, luazuaroiu2020consumers}, we ask systems to annotate their provided attributions with intent types: \textit{information seeking} (exploring products), \textit{alternative evaluation} (comparing options), and \textit{purchase decision} (finalizing selections), to identify which behaviors contribute most to personalization. 
Figure~\ref{fig:action_analysis} reveals that search-augmented LLMs primarily engage in alternative evaluation, with Claude (0.80) and GPT (0.74) investing the most effort in comparing options—correlating with their strongest performance scores.
We attribute this tendency to their inherent interaction architecture with search modules, where they are provided with pre-retrieved search results and product links, encouraging them to focus on comparing alternatives.

\medskip
\noindent\textbf{Exploration enhances effective personalized product curation.} Interestingly, we observe that the correlation between alternative evaluation and personalization becomes less pronounced among web agents, which exhibit higher personalization performance when measured on successful cases (in Table~\ref{tab:main_tasks_full}).
Moreover, web agents engage in substantially more information-seeking behaviors than search-augmented LLMs (in Figure~\ref{fig:action_analysis}), suggesting that exploratory information seeking contributes to effective personalized product curation.
Overall, balancing exploration with alternative comparison offers a promising direction for future work.

\medskip
\noindent\textbf{Error Analysis.}
To further understand the challenges faced by current agentic systems and provide practical directions for improvements, we conduct a comprehensive error analysis.
Detailed illustrations are in Appendix~\ref{appendix:error_analysis}. 
We find that hallucination remains a critical bottleneck, often leading to unsupported product curations.
Also, autonomous web agents are frequently distracted by promotional content, leading to incorrect product information.

\section{Conclusion}
\label{sec:conclusion}
In this work, we introduce \proposed, the first benchmark for evaluating agentic systems on personalized product curation in open web environments. 
By integrating realistic shopping intents, diverse user profiles, and checklist-driven personalization evaluation, \proposed enables rigorous and verifiable assessment of how well agents align with individual user needs.
Through extensive experiments, we reveal that current agentic systems struggle with personalized product curation.
Our work establishes \proposed as a foundation for developing and evaluating user-side shopping agents capable of personalized product curation.

\section*{Acknowledgments}
\label{sec:acknowledge}
This work was supported by the IITP grants funded by the Korea government (MSIT) (No. RS-2020-II201361; RS-2024-00457882, AI Research Hub Project).

\bibliographystyle{ACM-Reference-Format}
\balance
\bibliography{reference}

@article{peng2021does,
  title={How does information overload affect consumers’ online decision process? An event-related potentials study},
  author={Peng, Minjing and Xu, Zhicheng and Huang, Haiyang},
  journal={Frontiers in Neuroscience},
  volume={15},
  pages={695852},
  year={2021},
  publisher={Frontiers Media SA}
}

@article{lv2022impact,
  title={The impact of information overload of e-commerce platform on consumer return intention: Considering the moderating role of perceived environmental effectiveness},
  author={Lv, Jun and Liu, Xuan},
  journal={International Journal of Environmental Research and Public Health},
  volume={19},
  number={13},
  pages={8060},
  year={2022},
  publisher={MDPI}
}

@article{zhang2023examining,
  title={Examining the influence of information overload on consumers’ purchase in live streaming: A heuristic-systematic model perspective},
  author={Zhang, Guihua and Cao, Junwei and Liu, Dong},
  journal={Plos one},
  volume={18},
  number={8},
  pages={e0284466},
  year={2023},
  publisher={Public Library of Science San Francisco, CA USA}
}

@inproceedings{su2018user,
  title={User intent, behaviour, and perceived satisfaction in product search},
  author={Su, Ning and He, Jiyin and Liu, Yiqun and Zhang, Min and Ma, Shaoping},
  booktitle={Proceedings of the Eleventh ACM International Conference on Web Search and Data Mining},
  pages={547--555},
  year={2018}
}

@article{duan2013supporting,
  title={Supporting keyword search in product database: a probabilistic approach},
  author={Duan, Huizhong and Zhai, ChengXiang and Cheng, Jinxing and Kumar, Rohit},
  journal={Proceedings of the VLDB Endowment},
  volume={6},
  number={14},
  pages={1786--1797},
  year={2013},
  publisher={VLDB Endowment}
}

@inproceedings{van2016learning,
  title={Learning latent vector spaces for product search},
  author={Van Gysel, Christophe and de Rijke, Maarten and Kanoulas, Evangelos},
  booktitle={Proceedings of the 25th ACM international on conference on information and knowledge management},
  pages={165--174},
  year={2016}
}

@inproceedings{zhang2019neural,
  title={Neural IR meets graph embedding: A ranking model for product search},
  author={Zhang, Yuan and Wang, Dong and Zhang, Yan},
  booktitle={The World Wide Web Conference},
  pages={2390--2400},
  year={2019}
}

@inproceedings{dai2023contrastive,
  title={Contrastive learning for user sequence representation in personalized product search},
  author={Dai, Shitong and Liu, Jiongnan and Dou, Zhicheng and Wang, Haonan and Liu, Lin and Long, Bo and Wen, Ji-Rong},
  booktitle={Proceedings of the 29th ACM SIGKDD Conference on Knowledge Discovery and Data Mining},
  pages={380--389},
  year={2023}
}

@inproceedings{xu2024optimizing,
  title={Optimizing E-commerce Search: Toward a Generalizable and Rank-Consistent Pre-Ranking Model},
  author={Xu, Enqiang and Qiu, Yiming and Bai, Junyang and Zhang, Ping and Miao, Dadong and Wang, Songlin and Tang, Guoyu and Liu, Lin and Li, MingMing},
  booktitle={Proceedings of the 47th International ACM SIGIR Conference on Research and Development in Information Retrieval},
  pages={2875--2879},
  year={2024}
}

@article{darley2010toward,
  title={Toward an integrated framework for online consumer behavior and decision making process: A review},
  author={Darley, William K and Blankson, Charles and Luethge, Denise J},
  journal={Psychology \& marketing},
  volume={27},
  number={2},
  pages={94--116},
  year={2010},
  publisher={Wiley Online Library}
}

@article{athukorala2016exploratory,
  title={Is exploratory search different? A comparison of information search behavior for exploratory and lookup tasks},
  author={Athukorala, Kumaripaba and G{\l}owacka, Dorota and Jacucci, Giulio and Oulasvirta, Antti and Vreeken, Jilles},
  journal={Journal of the Association for Information Science and Technology},
  volume={67},
  number={11},
  pages={2635--2651},
  year={2016},
  publisher={Wiley Online Library}
}

@inproceedings{dammu2025shopping,
  title={A shopping agent for addressing subjective product needs},
  author={Dammu, Preetam Prabhu Srikar and Alonso, Omar and Poblete, Barbara},
  booktitle={Proceedings of the Eighteenth ACM International Conference on Web Search and Data Mining},
  pages={1032--1035},
  year={2025}
}

@inproceedings{kim2025user,
  title={User Orientations and Stage-Specific Behaviors in E-commerce Exploratory Search: A Formative Study},
  author={Kim, Eunhye and Choe, Kiroong and Yan, Guangjing and Kang, Mingyu},
  booktitle={Adjunct Proceedings of the 33rd ACM Conference on User Modeling, Adaptation and Personalization},
  pages={379--384},
  year={2025}
}

@inproceedings{sondhi2018taxonomy,
  title={A taxonomy of queries for e-commerce search},
  author={Sondhi, Parikshit and Sharma, Mohit and Kolari, Pranam and Zhai, ChengXiang},
  booktitle={The 41st International ACM SIGIR Conference on Research \& Development in Information Retrieval},
  pages={1245--1248},
  year={2018}
}

@article{jin2023amazon,
  title={Amazon-m2: A multilingual multi-locale shopping session dataset for recommendation and text generation},
  author={Jin, Wei and Mao, Haitao and Li, Zheng and Jiang, Haoming and Luo, Chen and Wen, Hongzhi and Han, Haoyu and Lu, Hanqing and Wang, Zhengyang and Li, Ruirui and others},
  journal={Advances in Neural Information Processing Systems},
  volume={36},
  pages={8006--8026},
  year={2023}
}

@inproceedings{chen2025chineseecomqa,
  title={Chineseecomqa: A scalable e-commerce concept evaluation benchmark for large language models},
  author={Chen, Haibin and Lv, Kangtao and Hu, Chengwei and Li, Yanshi and Yuan, Yujin and He, Yancheng and Zhang, Xingyao and Liu, Langming and Liu, Shilei and Su, Wenbo and others},
  booktitle={Proceedings of the 31st ACM SIGKDD Conference on Knowledge Discovery and Data Mining V. 2},
  pages={5311--5321},
  year={2025}
}

@article{yao2022webshop,
  title={Webshop: Towards scalable real-world web interaction with grounded language agents},
  author={Yao, Shunyu and Chen, Howard and Yang, John and Narasimhan, Karthik},
  journal={Advances in Neural Information Processing Systems},
  volume={35},
  pages={20744--20757},
  year={2022}
}

@inproceedings{liu2023webglm,
  title={WebGLM: towards an efficient web-enhanced question answering system with human preferences},
  author={Liu, Xiao and Lai, Hanyu and Yu, Hao and Xu, Yifan and Zeng, Aohan and Du, Zhengxiao and Zhang, Peng and Dong, Yuxiao and Tang, Jie},
  booktitle={Proceedings of the 29th ACM SIGKDD conference on knowledge discovery and data mining},
  pages={4549--4560},
  year={2023}
}

@article{deng2023mind2web,
  title={Mind2web: Towards a generalist agent for the web},
  author={Deng, Xiang and Gu, Yu and Zheng, Boyuan and Chen, Shijie and Stevens, Sam and Wang, Boshi and Sun, Huan and Su, Yu},
  journal={Advances in Neural Information Processing Systems},
  volume={36},
  pages={28091--28114},
  year={2023}
}

@inproceedings{zhouwebarena,
  title={WebArena: A Realistic Web Environment for Building Autonomous Agents},
  author={Zhou, Shuyan and Xu, Frank F and Zhu, Hao and Zhou, Xuhui and Lo, Robert and Sridhar, Abishek and Cheng, Xianyi and Ou, Tianyue and Bisk, Yonatan and Fried, Daniel and others},
  booktitle={The Twelfth International Conference on Learning Representations}
}

@inproceedings{pan2024webcanvas,
  title={WebCanvas: Benchmarking Web Agents in Online Environments},
  author={Pan, Yichen and Kong, Dehan and Zhou, Sida and Cui, Cheng and Leng, Yifei and Jiang, Bing and Liu, Hangyu and Shang, Yanyi and Zhou, Shuyan and Wu, Tongshuang and others},
  booktitle={Agentic Markets Workshop at ICML 2024}
}

@inproceedings{he2024webvoyager,
  title={WebVoyager: Building an End-to-End Web Agent with Large Multimodal Models},
  author={He, Hongliang and Yao, Wenlin and Ma, Kaixin and Yu, Wenhao and Dai, Yong and Zhang, Hongming and Lan, Zhenzhong and Yu, Dong},
  booktitle={Proceedings of the 62nd Annual Meeting of the Association for Computational Linguistics (Volume 1: Long Papers)},
  pages={6864--6890},
  year={2024}
}

@inproceedings{koh2024visualwebarena,
  title={VisualWebArena: Evaluating Multimodal Agents on Realistic Visual Web Tasks},
  author={Koh, Jing Yu and Lo, Robert and Jang, Lawrence and Duvvur, Vikram and Lim, Ming and Huang, Po-Yu and Neubig, Graham and Zhou, Shuyan and Salakhutdinov, Russ and Fried, Daniel},
  booktitle={Proceedings of the 62nd Annual Meeting of the Association for Computational Linguistics (Volume 1: Long Papers)},
  pages={881--905},
  year={2024}
}

@article{chen2024chatshop,
  title={Chatshop: Interactive information seeking with language agents},
  author={Chen, Sanxing and Wiseman, Sam and Dhingra, Bhuwan},
  journal={arXiv preprint arXiv:2404.09911},
  year={2024}
}

@inproceedings{lu2024weblinx,
  title={WEBLINX: real-world website navigation with multi-turn dialogue},
  author={L{\`u}, Xing Han and Kasner, Zden{\v{e}}k and Reddy, Siva},
  booktitle={Proceedings of the 41st International Conference on Machine Learning},
  pages={33007--33056},
  year={2024}
}

@article{jin2024shopping,
  title={Shopping mmlu: A massive multi-task online shopping benchmark for large language models},
  author={Jin, Yilun and Li, Zheng and Zhang, Chenwei and Cao, Tianyu and Gao, Yifan and Jayarao, Pratik and Li, Mao and Liu, Xin and Sarkhel, Ritesh and Tang, Xianfeng and others},
  journal={Advances in Neural Information Processing Systems},
  volume={37},
  pages={18062--18089},
  year={2024}
}

@inproceedings{cai2025large,
  title={Large language models empowered personalized web agents},
  author={Cai, Hongru and Li, Yongqi and Wang, Wenjie and Zhu, Fengbin and Shen, Xiaoyu and Li, Wenjie and Chua, Tat-Seng},
  booktitle={Proceedings of the ACM on Web Conference 2025},
  pages={198--215},
  year={2025}
}

@article{gou2025mind2web,
  title={Mind2Web 2: Evaluating Agentic Search with Agent-as-a-Judge},
  author={Gou, Boyu and Huang, Zanming and Ning, Yuting and Gu, Yu and Lin, Michael and Qi, Weijian and Kopanev, Andrei and Yu, Botao and Guti{\'e}rrez, Bernal Jim{\'e}nez and Shu, Yiheng and others},
  journal={arXiv preprint arXiv:2506.21506},
  year={2025}
}

@article{lyu2025deepshop,
  title={DeepShop: A Benchmark for Deep Research Shopping Agents},
  author={Lyu, Yougang and Zhang, Xiaoyu and Yan, Lingyong and de Rijke, Maarten and Ren, Zhaochun and Chen, Xiuying},
  journal={arXiv preprint arXiv:2506.02839},
  year={2025}
}

@article{wang2025shoppingbench,
  title={ShoppingBench: A Real-World Intent-Grounded Shopping Benchmark for LLM-based Agents},
  author={Wang, Jiangyuan and Xiao, Kejun and Sun, Qi and Zhao, Huaipeng and Luo, Tao and Zhang, Jiandong and Zeng, Xiaoyi},
  journal={arXiv preprint arXiv:2508.04266},
  year={2025}
}

@article{peeters2025webmall,
  title={WebMall--A Multi-Shop Benchmark for Evaluating Web Agents},
  author={Peeters, Ralph and Steiner, Aaron and Schwarz, Luca and Caspary, Julian Yuya and Bizer, Christian},
  journal={arXiv preprint arXiv:2508.13024},
  year={2025}
}

@inproceedings{xue2025an,
title={An Illusion of Progress? Assessing the Current State of Web Agents},
author={Tianci Xue and Weijian Qi and Tianneng Shi and Chan Hee Song and Boyu Gou and Dawn Song and Huan Sun and Yu Su},
booktitle={Second Conference on Language Modeling},
year={2025},
url={https://openreview.net/forum?id=6jZi4HSs6o}
}

@article{zhang2025functionality,
  title={A Functionality-Grounded Benchmark for Evaluating Web Agents in E-commerce Domains},
  author={Zhang, Xianren and Prasad, Shreyas and Wang, Di and Zeng, Qiuhai and Wang, Suhang and Yan, Wenbo and Hans, Mat},
  journal={arXiv preprint arXiv:2508.15832},
  year={2025}
}

@article{hou2024bridging,
  title={Bridging Language and Items for Retrieval and Recommendation},
  author={Hou, Yupeng and Li, Jiacheng and He, Zhankui and Yan, An and Chen, Xiusi and McAuley, Julian},
  journal={arXiv preprint arXiv:2403.03952},
  year={2024}
}

@inproceedings{
zheng2024gptvision,
title={{GPT}-4V(ision) is a Generalist Web Agent, if Grounded},
author={Boyuan Zheng and Boyu Gou and Jihyung Kil and Huan Sun and Yu Su},
booktitle={Forty-first International Conference on Machine Learning},
year={2024},
url={https://openreview.net/forum?id=piecKJ2DlB}
}

@inproceedings{
gou2025navigating,
title={Navigating the Digital World as Humans Do: Universal Visual Grounding for {GUI} Agents},
author={Boyu Gou and Ruohan Wang and Boyuan Zheng and Yanan Xie and Cheng Chang and Yiheng Shu and Huan Sun and Yu Su},
booktitle={The Thirteenth International Conference on Learning Representations},
year={2025},
url={https://openreview.net/forum?id=kxnoqaisCT}
}

@misc{anthropic2024claudeuse,
  author       = {Anthropic},
  title        = {Claude computer use},
  year         = {2024},
  howpublished = {\url{https://www.anthropic.com/news/3-5-models-and-computer-use}},
  note         = {Accessed: 2025-05-08}
}

@misc{perplexity2024,
  author       = {{Perplexity AI}},
  title        = {Perplexity AI},
  year         = {2024},
  howpublished = {\url{https://www.perplexity.ai/}},
  note         = {Accessed: 2025-05-08}
}

@article{comanici2025gemini,
  title={Gemini 2.5: Pushing the frontier with advanced reasoning, multimodality, long context, and next generation agentic capabilities},
  author={Comanici, Gheorghe and Bieber, Eric and Schaekermann, Mike and Pasupat, Ice and Sachdeva, Noveen and Dhillon, Inderjit and Blistein, Marcel and Ram, Ori and Zhang, Dan and Rosen, Evan and others},
  journal={arXiv preprint arXiv:2507.06261},
  year={2025}
}

@article{openai2023gpt4,
  title        = {GPT-4 Technical Report},
  author       = {OpenAI},
  year         = {2023},
  journal      = {arXiv preprint arXiv:2303.08774},
  url          = {https://api.semanticscholar.org/CorpusID:257532815}
}

@techreport{openai2025operator,
  author      = {OpenAI},
  title       = {Operator System Card},
  institution = {OpenAI},
  year        = {2025},
  month       = jan,
  note        = {Technical report},
  url         = {https://cdn.openai.com/operator_system_card.pdf}
}

@misc{openai2024chatgptsearch,
  author       = {OpenAI},
  title        = {Introducing ChatGPT Search},
  year         = {2024},
  url          = {https://openai.com/index/introducing-chatgpt-search/}
}

@misc{anthropic2024websearch,
  author= {Anthropic},
  title= {Claude can now search the web},
  year= {2024},
  url= {https://www.anthropic.com/news/web-search}
}

@inproceedings{abuelsaad2024agent,
  title={Agent-E: From Autonomous Web Navigation to Foundational Design Principles in Agentic Systems},
  author={Abuelsaad, Tamer and Akkil, Deepak and Dey, Prasenjit and Jagmohan, Ashish and Vempaty, Aditya and Kokku, Ravi},
  booktitle={NeurIPS 2024 Workshop on Open-World Agents}
}

@misc{muller2024browseruse,
  author = {M{\"a}gnus M{\"u}ller and Gregor {\v{Z}}uni{\v{c}}},
  title  = {{Browser Use} = {State of the Art Web Agent}},
  year   = {2024},
  url    = {https://browser-use.com/posts/sota-technical-report},
}

@article{song2024beyond,
  title={Beyond Browsing: API-Based Web Agents},
  author={Song, Yueqi and Xu, Frank F and Zhou, Shuyan and Neubig, Graham},
  journal={CoRR},
  year={2024}
}

@article{nakano2021webgpt,
  title={Webgpt: Browser-assisted question-answering with human feedback},
  author={Nakano, Reiichiro and Hilton, Jacob and Balaji, Suchir and Wu, Jeff and Ouyang, Long and Kim, Christina and Hesse, Christopher and Jain, Shantanu and Kosaraju, Vineet and Saunders, William and others},
  journal={arXiv preprint arXiv:2112.09332},
  year={2021}
}

@article{panwar2019consumer,
  title={Consumer decision making process models and their applications to market strategy},
  author={Panwar, Diksha and Anand, Swati and Ali, Farmaan and Singal, Kanika},
  journal={International Management Review},
  volume={15},
  number={1},
  pages={36--44},
  year={2019},
  publisher={American Scholars Press, Inc.}
}

@article{gur2023real,
  title={A real-world webagent with planning, long context understanding, and program synthesis},
  author={Gur, Izzeddin and Furuta, Hiroki and Huang, Austin and Safdari, Mustafa and Matsuo, Yutaka and Eck, Douglas and Faust, Aleksandra},
  journal={arXiv preprint arXiv:2307.12856},
  year={2023}
}

@article{koh2024tree,
  title={Tree search for language model agents},
  author={Koh, Jing Yu and McAleer, Stephen and Fried, Daniel and Salakhutdinov, Ruslan},
  journal={arXiv preprint arXiv:2407.01476},
  year={2024}
}

@article{furuta2023multimodal,
  title={Multimodal web navigation with instruction-finetuned foundation models},
  author={Furuta, Hiroki and Lee, Kuang-Huei and Nachum, Ofir and Matsuo, Yutaka and Faust, Aleksandra and Gu, Shixiang Shane and Gur, Izzeddin},
  journal={arXiv preprint arXiv:2305.11854},
  year={2023}
}

@article{gu2024your,
  title={Is your llm secretly a world model of the internet? model-based planning for web agents},
  author={Gu, Yu and Zhang, Kai and Ning, Yuting and Zheng, Boyuan and Gou, Boyu and Xue, Tianci and Chang, Cheng and Srivastava, Sanjari and Xie, Yanan and Qi, Peng and others},
  journal={arXiv preprint arXiv:2411.06559},
  year={2024}
}

@inproceedings{mialon2023gaia,
  title={Gaia: a benchmark for general ai assistants},
  author={Mialon, Gr{\'e}goire and Fourrier, Cl{\'e}mentine and Wolf, Thomas and LeCun, Yann and Scialom, Thomas},
  booktitle={The Twelfth International Conference on Learning Representations},
  year={2023}
}

@article{zheng2023judging,
  title={Judging llm-as-a-judge with mt-bench and chatbot arena},
  author={Zheng, Lianmin and Chiang, Wei-Lin and Sheng, Ying and Zhuang, Siyuan and Wu, Zhanghao and Zhuang, Yonghao and Lin, Zi and Li, Zhuohan and Li, Dacheng and Xing, Eric and others},
  journal={Advances in neural information processing systems},
  volume={36},
  pages={46595--46623},
  year={2023}
}

@article{kim2024stop,
  title={Stop playing the guessing game! target-free user simulation for evaluating conversational recommender systems},
  author={Kim, Sunghwan and Kim, Tongyoung and Seo, Kwangwook and Yeo, Jinyoung and Lee, Dongha},
  journal={arXiv preprint arXiv:2411.16160},
  year={2024}
}

@article{kim2025towards,
  title={Towards Personalized Conversational Sales Agents: Contextual User Profiling for Strategic Action},
  author={Kim, Tongyoung and Lee, Jeongeun and Yoon, Soojin and Kim, Sunghwan and Lee, Dongha},
  journal={arXiv preprint arXiv:2504.08754},
  year={2025}
}

@article{kim2025bespoke,
  title={BESPOKE: Benchmark for Search-Augmented Large Language Model Personalization via Diagnostic Feedback},
  author={Kim, Hyunseo and Lee, Sangam and Seo, Kwangwook and Lee, Dongha},
  journal={arXiv preprint arXiv:2509.21106},
  year={2025}
}

@article{miroyan2025search,
  title={Search Arena: Analyzing Search-Augmented LLMs},
  author={Miroyan, Mihran and Wu, Tsung-Han and King, Logan and Li, Tianle and Pan, Jiayi and Hu, Xinyan and Chiang, Wei-Lin and Angelopoulos, Anastasios N and Darrell, Trevor and Norouzi, Narges and others},
  journal={arXiv preprint arXiv:2506.05334},
  year={2025}
}

@article{luazuaroiu2020consumers,
  title={Consumers’ decision-making process on social commerce platforms: Online trust, perceived risk, and purchase intentions},
  author={L{\u{a}}z{\u{a}}roiu, George and Neguri{\c{t}}{\u{a}}, Octav and Grecu, Iulia and Grecu, Gheorghe and Mitran, Paula Cornelia},
  journal={Frontiers in psychology},
  volume={11},
  pages={890},
  year={2020},
  publisher={Frontiers Media SA}
}

@article{heo2025can,
  title={Can Large Language Models be Effective Online Opinion Miners?},
  author={Heo, Ryang and Seo, Yongsik and Lee, Junseong and Lee, Dongha},
  journal={arXiv preprint arXiv:2505.15695},
  year={2025}
}

@inproceedings{tian2025mmina,
  title={Mmina: Benchmarking multihop multimodal internet agents},
  author={Tian, Shulin and Zhang, Ziniu and Chen, Liang-Yu and Liu, Ziwei},
  booktitle={Findings of the Association for Computational Linguistics: ACL 2025},
  pages={13682--13697},
  year={2025}
}

\appendix

\setlength{\tabcolsep}{2pt}
\renewcommand{\arraystretch}{1.1}
\begin{table*}[t]
\caption{Performance comparison of agentic systems on AgenticShop, extended to cases where all product pages are valid.}
\centering
\scriptsize
\resizebox{0.89\textwidth}{!}{
\begin{tabular}{p{2.5cm} *{14}{c}}
\toprule
\multirow{4}{*}{\textbf{Agent}}
& \multicolumn{6}{c}{\textbf{TF}} 
& \multicolumn{6}{c}{\textbf{AS}} 
& \multicolumn{2}{c}{\textbf{OE}} \\
\cmidrule(lr){2-7} \cmidrule(lr){8-13} \cmidrule(lr){14-15}
& \multicolumn{2}{c}{\textbf{Exp. Title}} 
& \multicolumn{2}{c}{\textbf{Attr. Spec.}} 
& \multicolumn{2}{c}{\textbf{Brand Cat.}} 
& \multicolumn{2}{c}{\textbf{High Inv.}} 
& \multicolumn{2}{c}{\textbf{Low Inv.}} 
& \multicolumn{2}{c}{\textbf{Aesthe. Driv.}} 
& \multicolumn{2}{c}{\textbf{Cas. Brow.}} \\
\cmidrule(lr){2-3} \cmidrule(lr){4-5} \cmidrule(lr){6-7} 
\cmidrule(lr){8-9} \cmidrule(lr){10-11} \cmidrule(lr){12-13} \cmidrule(lr){14-15}
& \textbf{All} & \textbf{Success} 
& \textbf{All} & \textbf{Success} 
& \textbf{All} & \textbf{Success} 
& \textbf{All} & \textbf{Success} 
& \textbf{All} & \textbf{Success} 
& \textbf{All} & \textbf{Success} 
& \textbf{All} & \textbf{Success} \\
\midrule 
\multicolumn{15}{l}{\textbf{\textit{Search-augmented LLMs}}} \\
\gpt         & \underline{37.81} & \underline{43.08} & \underline{35.35} & \underline{39.31} & \underline{34.79} & \underline{39.86} & \textbf{30.13} & 37.64 & \textbf{29.47} & \underline{39.31} & \textbf{27.55} & \textbf{38.95} & 14.51 & 22.41 \\
\claude      & \textbf{37.93} & \textbf{44.52} & \textbf{36.44} & \textbf{42.76} & \textbf{34.96} & \textbf{42.64} & \underline{29.66} & \underline{39.20} & \underline{29.08} & \textbf{40.02} & \underline{20.58} & \underline{36.15} & \textbf{21.20} & \textbf{28.65} \\
\gemini      & 22.29 & 39.24 & 21.52 & 36.40 & 23.60 & 34.97 & 18.04 & 33.19 & 19.72 & 30.70 & 16.62 & 35.61 & \underline{16.06} & \underline{26.10} \\
\pplx        & 22.85 & 38.65 & 22.04 & 38.04 & 21.03 & 37.42 & 19.95 & \textbf{42.90} & 22.57 & 36.98 & 14.36 & 32.75 & 14.36 & 25.03 \\
\midrule
\multicolumn{15}{l}{\textbf{\textit{Autonomous Web Agents}}} \\
Agent-E      & 19.75 & \textbf{54.02} & 20.65 & \textbf{46.39} & 14.85 & \textbf{47.98} & 16.15 & \textbf{43.00} & 21.76 & \textbf{51.23} & \underline{19.98} & \textbf{44.92} & 13.56 & 17.43 \\
SeeAct       & \textbf{37.13} & 44.14 & \textbf{36.32} & 45.13 & \textbf{34.92} & 43.40 & \underline{25.50} & 35.80 & \textbf{27.47} & \underline{42.39} & \textbf{20.80} & \underline{35.60} & \textbf{20.79} & \textbf{28.89} \\
Web Voyager  & \underline{34.25} & 42.57 & 30.34 & 45.20 & 27.22 & \underline{44.02} & \textbf{25.87} & \underline{36.37} & \underline{25.68} & 32.52 & 18.76 & 32.28 & 15.57 & \underline{26.91} \\
Browser Use  & 31.48 & \underline{45.81} & \underline{34.52} & \underline{45.38} & \underline{30.58} & 42.31 & 22.61 & 34.77 & 23.45 & 41.21 & 16.24 & 34.87 & \underline{18.41} & 26.57 \\
\bottomrule
\end{tabular}
}
\label{tab:main_tasks_full}
\end{table*}
\begin{table}[ht]
\centering
\caption{{The statistics of \proposed. \#Product denotes the distinct product type associated with each user profile. Avg. $\boldsymbol{{|\mathcal{C}|}}$ denotes the average number of checklist items per user.}}
\label{tab:bench_stats}
\begin{tabular}{cccccc}
\toprule
\textbf{Intent} & \textbf{Scenario} & \textbf{Domain} & \textbf{\#Users} & \textbf{\#Products} & \textbf{Avg.} $\boldsymbol{{|\mathcal{C}|}}$ \\ 
\midrule
\multirow{3}{*}{TF} 
   & Exp. Title & \textit{Food} & 50 & 50 & 19.3  \\
   & Attr. Spec. & \textit{Food} & 50 & 50 & 19.3 \\ 
   & Brand Cat. & \textit{Food} & 50 & 50 & 19.3  \\
\midrule
\multirow{3}{*}{AS} 
   & Low Inv. & \textit{Home} & 50 & 50 & 19.6 \\
   & High Inv.  & \textit{Electronics} & 50 & 50 & 20.0  \\
   & Aesth. Dri. & \textit{Fashion} & 50 & 50 & 19.4  \\
\midrule
\multirow{1}{*}{OE}   
   & Cas. Brow & \textit{All} & 50 & - & 19.5 \\
\bottomrule
\end{tabular}
\end{table}
\begin{figure}[ht]
    \centering
    \includegraphics[width=0.99\linewidth]{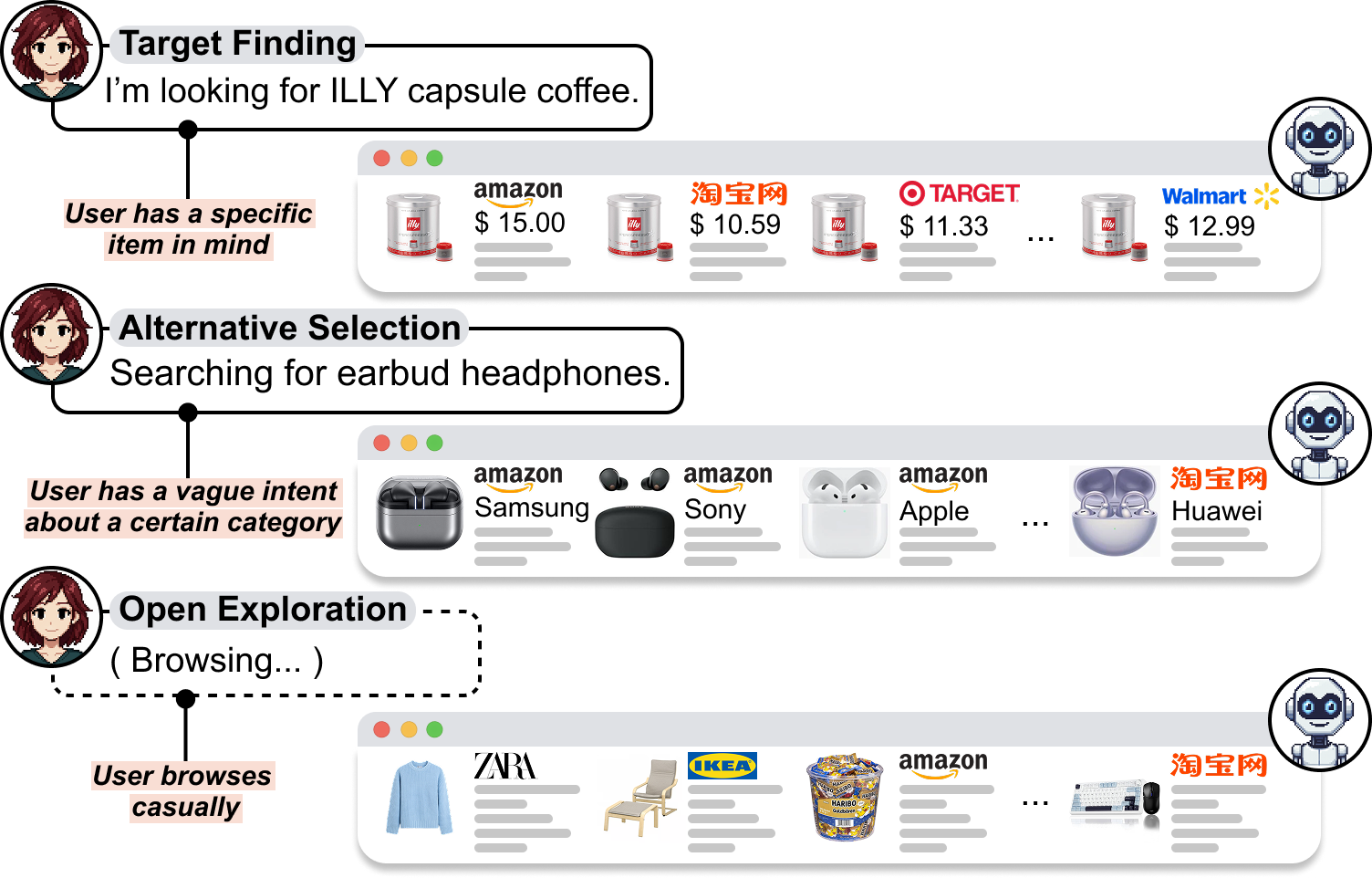}
    \caption{Three user intents in real-world web shopping.}
    \label{fig:shopping_scenario_fig}
\end{figure}

\section{Benchmark Details}
\label{appendix:dataset}
Table~\ref{tab:bench_stats} provides an overview of the composition and statistics of \proposed.
We construct 50 user profiles for each shopping scenario, resulting in a total of 350 personalized tasks, each accompanied by a tailored checklist capturing user-specific preferences. 
Each task specifies a distinct product type that the user is seeking, leading to 50 unique products per scenario. 
An exception is made for OE (Open Exploration) tasks, in which users engage in casual browsing without predetermined product categories.

\section{Dimension-wise Evaluation Reliability Analysis}
\label{appendix:checklist_meta_eval}
To further validate the reliability of our checklist-driven evaluation framework, we provide additional human evaluation results across the six checklist dimensions.
As shown in Table~\ref{tab:human_eval_dimensions}, our approach shows substantial agreement with human evaluators, with both Cohen's $\kappa$ and Spearman's $\rho$ approximately 0.7 across all dimensions.

\section{Analysis of Successful Cases}
\label{appendix:main_tasks_full}
Table~\ref{tab:main_tasks_full} reports the results for cases where all curated product URLs were valid and accessible, ensuring that each agent response is grounded in a real product page.
As shown in the table, both search-augmented LLMs and autonomous web agents show higher personalization scores when failure cases are excluded, indicating the importance of robust task completion. 
The improvement is particularly large for autonomous web agents, surpassing search-augmented LLMs. 
We attribute this to their iterative self-evaluation mechanism, as web agents can dynamically explore more information and evaluate their state throughout multiple steps. 
We also find that they are more capable of recovering from mistakes.
For example, they may choose to return to a previous page if a product does not meet requirements. 
Nevertheless, this increased exploration horizon may lead to suboptimal personalization scores as agents struggle to synthesize information from too many sources.

\section{Error Analysis}
\label{appendix:error_analysis}
\begin{figure*}[t]
\centering

\subfigure[Search-augmented LLMs hallucinate Amazon product links by fabricating ID characters.]{
  \includegraphics[width=0.48\linewidth]{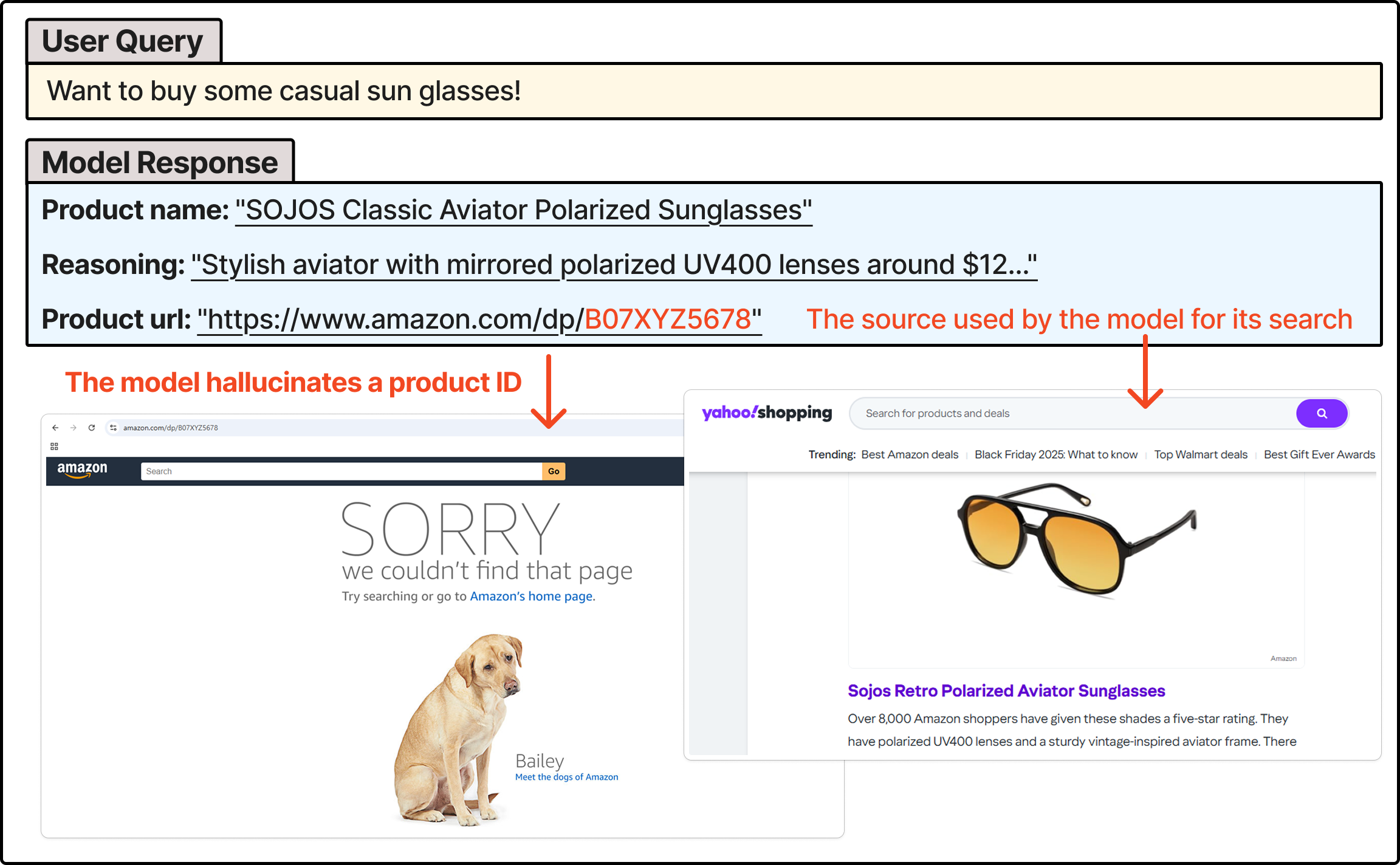}
  \label{fig:url_hallucination}
}
\hfill
\subfigure[Retrieval module fails to provide valid grounding URLs, yet the model hallucinates detailed explanations.]{
  \includegraphics[width=0.48\linewidth]{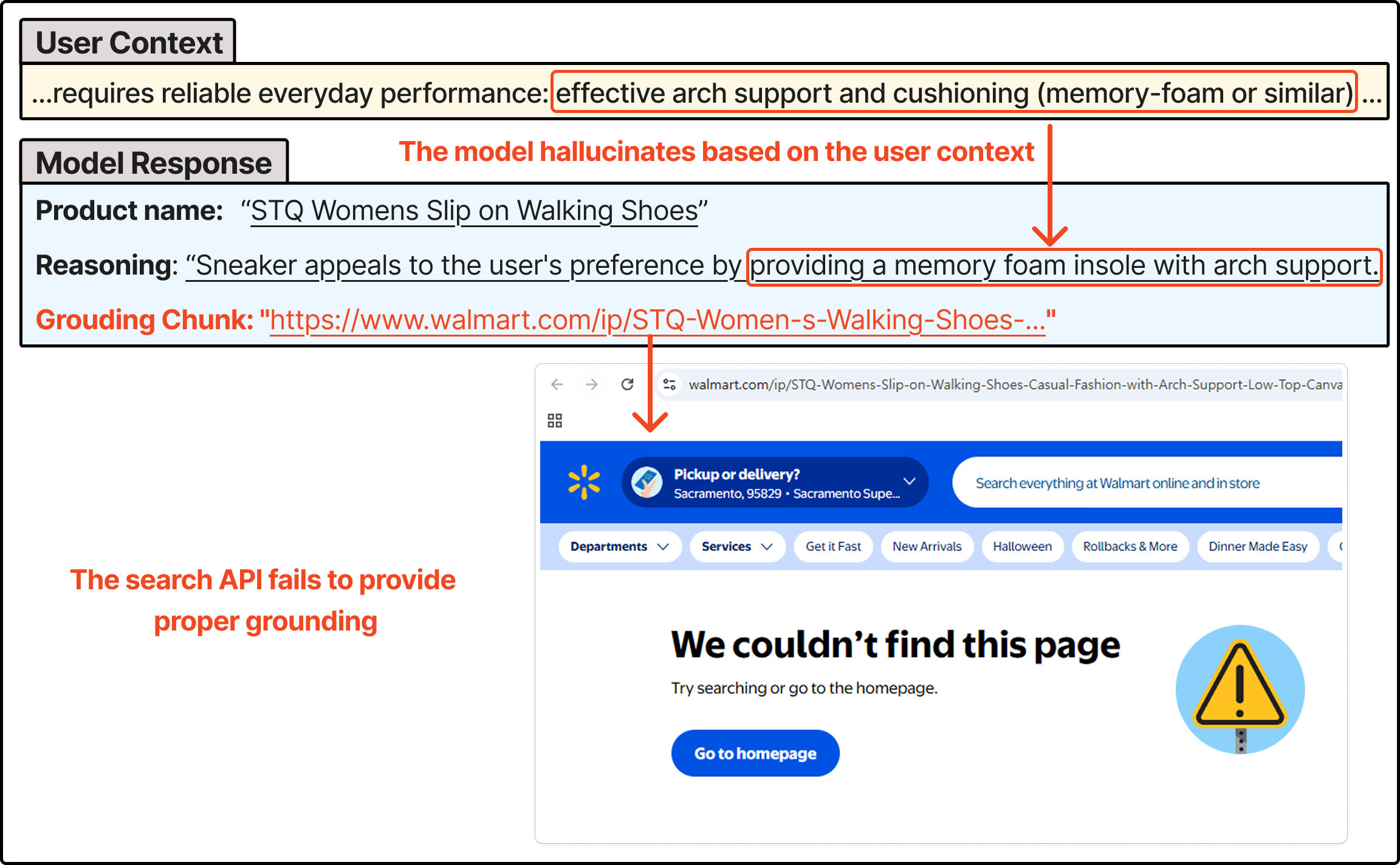}
  \label{fig:grounding_response_hallucination}
}

\vspace{0.4cm}

\subfigure[Query bias in search-augmented LLMs and over-complicated queries in autonomous web agents.]{
  \includegraphics[width=0.48\linewidth]{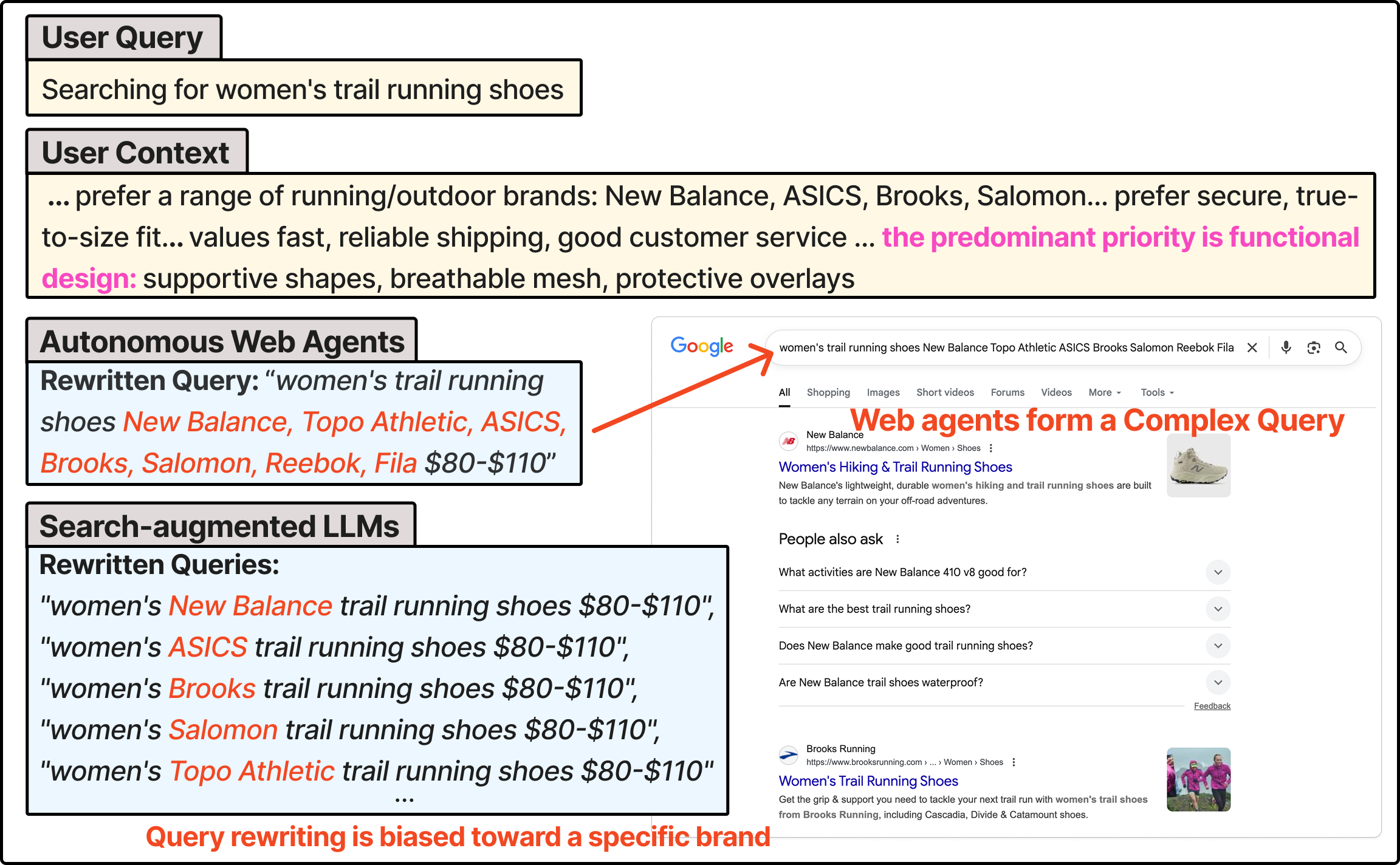}
  \label{fig:query_bias}
}
\hfill
\subfigure[Autonomous web agents prematurely terminate exploration without detailed product inspection.]{
  \includegraphics[width=0.48\linewidth]{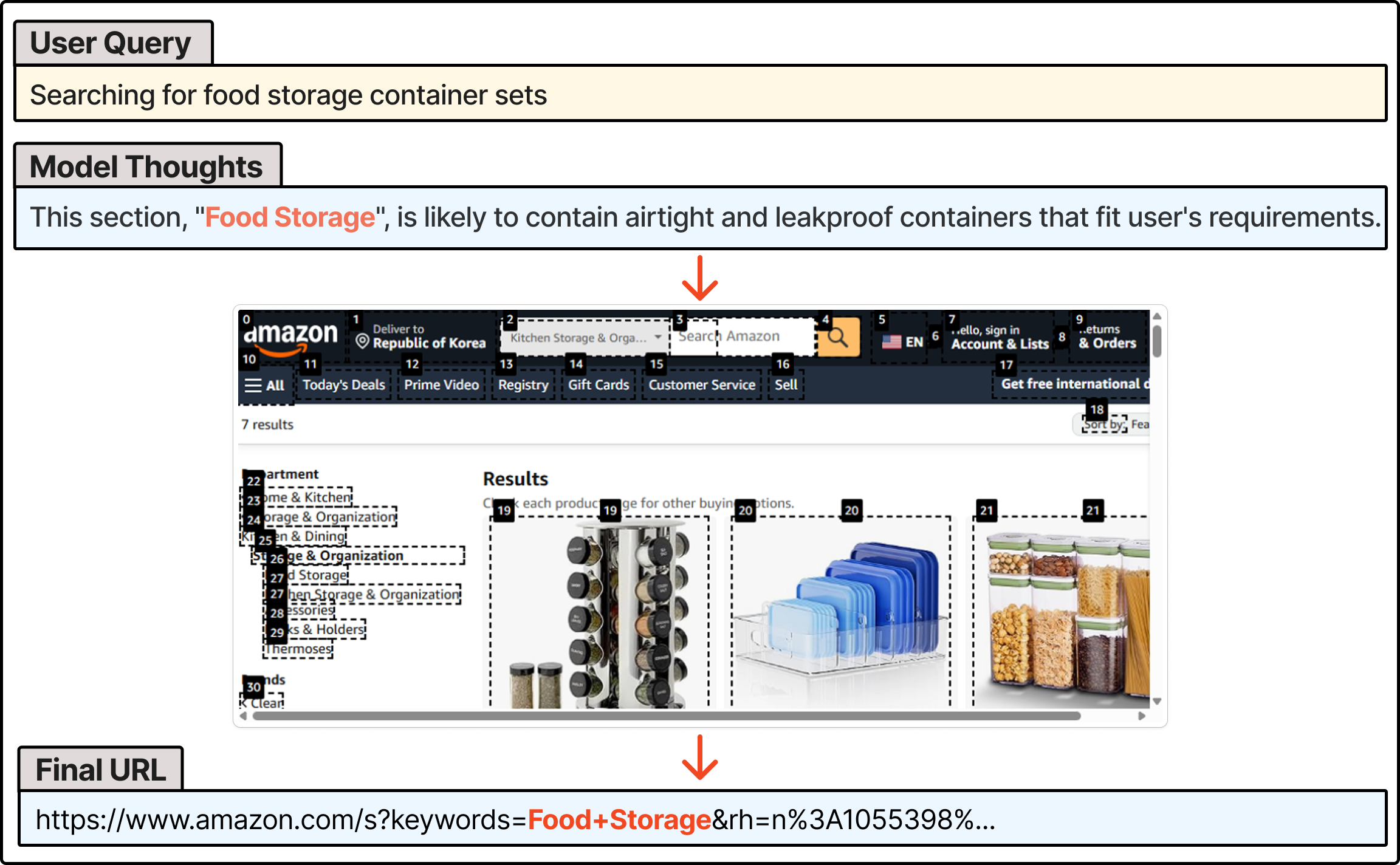}
  \label{fig:premature_termination}
}

\vspace{0.4cm}

\subfigure[Autonomous web agents misidentify product price due to distraction by nearby alternatives.]{
  \includegraphics[width=0.48\linewidth]{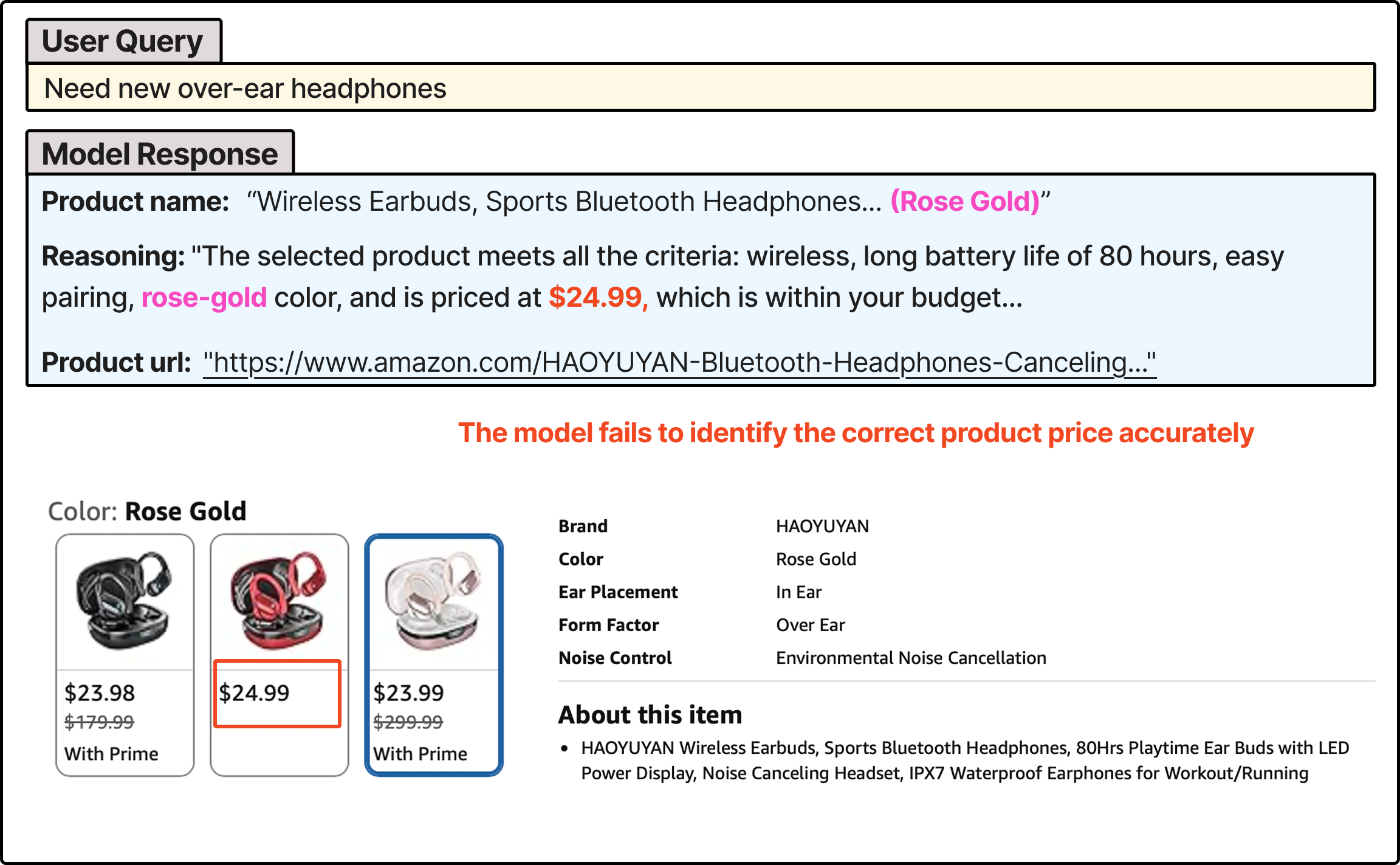}
  \label{fig:distraction}
}
\hfill
\subfigure[Autonomous web agents fail to adapt exploration strategy after encountering out-of-budget products.]{
  \includegraphics[width=0.48\linewidth]{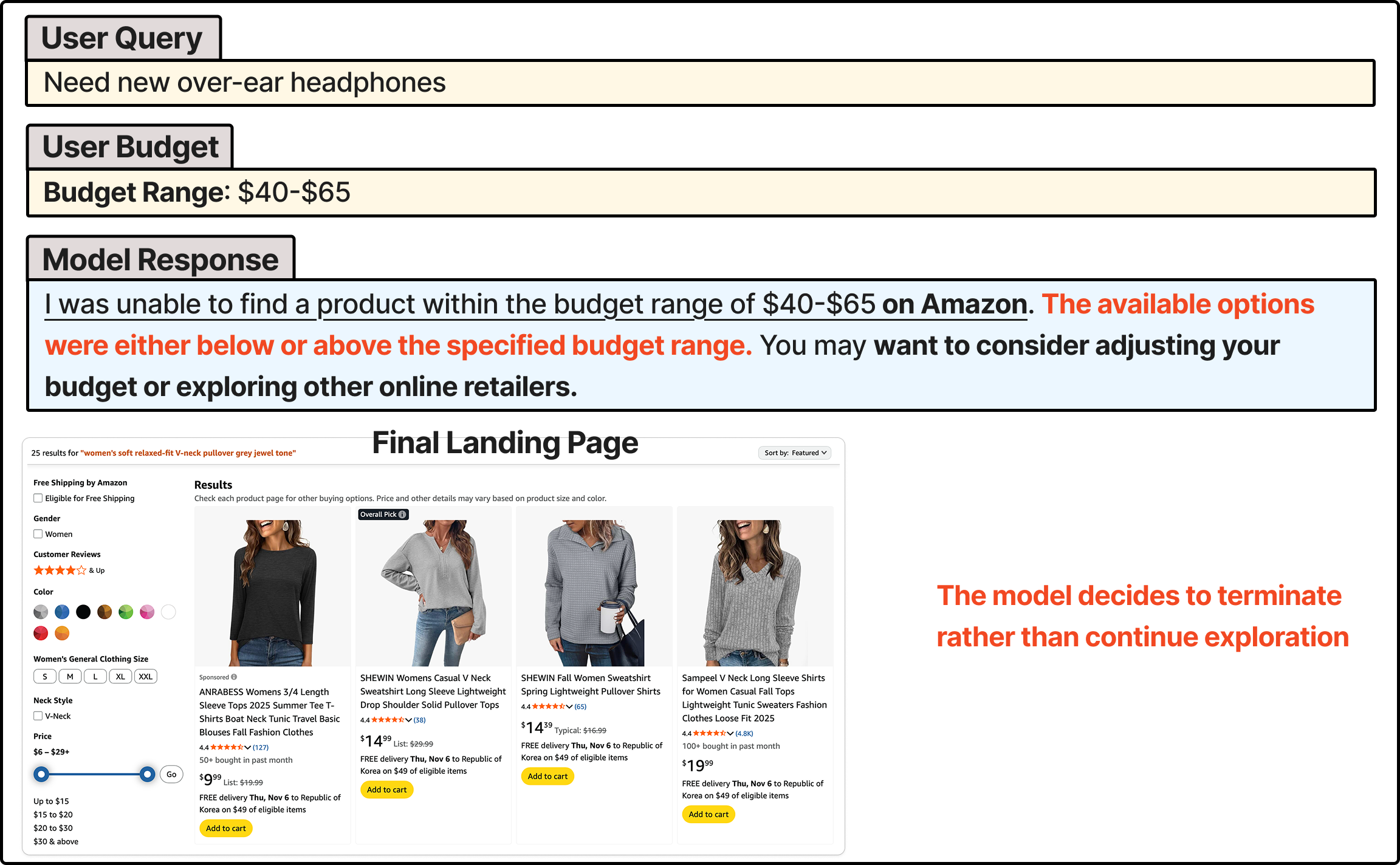}
  \label{fig:insufficient_adaptation}
}

\caption{Representative failure cases of search-augmented LLMs and autonomous web agents.}
\label{fig:error_cases}
\end{figure*}

We manually examine system logs, generated responses, and navigation traces to identify recurring failure patterns and underlying causes of agentic systems.
Detailed findings are as follows:

\medskip
\noindent\textbf{Search-augmented LLMs.} Through manual inspection of system outputs and search logs, we identify four common failure patterns:

\setlength{\leftmargini}{10pt}
\setlength{\itemsep}{2pt}
\begin{itemize}
    \item \textbf{\textit{URL hallucination}.} Models frequently generate malformed product links through character omission or fabrication. Common patterns include hallucinated ASINs on Amazon (Figure ~\ref{fig:url_hallucination}) or missing path segments between forward slashes. To investigate the cause of these errors, we inspect search logs to check if models mistakenly drop or modify characters when formatting URLs from search modules. However, examining Gemini's grounded citations in its structured outputs reveals that the search tool itself sometimes produces invalid URLs (Figure~\ref{fig:grounding_response_hallucination}), indicating the error may originates at the retrieval stage.
    
    \item \textit{\textbf{Response hallucination.}} We observe that models frequently generate detailed product explanations even when the grounding links are hallucinated (Figure~\ref{fig:grounding_response_hallucination}), by leveraging the user context provided in their prompt. We also observe that even when the product links are valid, models sometimes state incorrect attributes like price, claiming products are within budget even though the actual prices exceed the specified range. 
    
    \item \textit{\textbf{Biased query rewriting.}} We analyze the search queries internally used by search-augmented LLMs, as models like Gemini provide access to their search queries. As shown in Figure~\ref{fig:query_bias}, we observe that they often prioritize brand when rewriting queries, generating brand-focused queries despite user context emphasizing functional attributes. This suggests that search-augmented LLMs may overweight certain attributes in query generation.
\end{itemize}

\begin{table}[t]
\centering
\scriptsize
\caption{Correlation between LLM-as-a-judge and human judgments across six personalization dimensions.}
\resizebox{0.9\columnwidth}{!}{
\begin{tabular}{lccc}
\toprule
\textbf{Dimension} & \textbf{Cohen's $\kappa$}  & \textbf{Spearman's $\rho$} & \textbf{Inter-Human $\rho$} \\
\midrule
Brand Pref.      & 0.792 & 0.792 & 0.849 \\
Price Sens.      & 0.691 & 0.706 & 0.743 \\
Review Sens.     & 0.674 & 0.690 & 0.786 \\
Func. Req.       & 0.711 & 0.722 & 0.722 \\
Aesth. Pref.     & 0.731 & 0.741 & 0.738 \\
Purchase Pref.   & 0.704 & 0.737 & 0.750 \\
\midrule
\textbf{Overall} & \textbf{0.732} & \textbf{0.743} & \textbf{0.774} \\
\bottomrule
\end{tabular}       
}
\label{tab:human_eval_dimensions}
\end{table}
\medskip
\noindent\textbf{Autonomous web agents.} Web agents exhibit distinct failure modes stemming from their interactive nature:

\setlength{\leftmargini}{10pt}
\setlength{\itemsep}{2pt}
\begin{itemize}

    \item \textit{\textbf{URL hallucination.}} Similar to search-augmented LLMs, web agents also generate invalid product links when terminating tasks. We find that this may be attributed to lengthy input contexts during the response generation stage, as they often need to both reason about the next action based on observation and assess personalization simultaneously.
    
    \item \textit{\textbf{Overcomplicated search queries.}} We observe that autonomous web agents often rewrite user queries into overly complicated forms by concatenating multiple attributes into single searches (Figure~\ref{fig:query_bias}). These queries may perform poorly in open web environments for two reasons: (1) the open web lacks structured taxonomies found in shopping platforms, and (2) such complex queries return inconsistent search results. This potentially poses challenges in retrieving and integrating relevant information.
    
    \item \textit{\textbf{Premature termination.}} Autonomous web agents frequently stop exploration before reaching product pages, even though we explicitly instruct them to return final product links. They often terminate at product listing pages with a few alternatives available (Figure ~\ref{fig:premature_termination}), without conducting further examination of individual products details. 
    
    \item \textit{\textbf{Distraction by promotional content.}} We find cases where autonomous web agents incorrectly identify product attributes, distracted by promotional elements on the page. For example, they sometimes misidentify prices from alternatives (Figure ~\ref{fig:distraction}) instead of extracting information from the main product. This may cause incorrect budget evaluations and lead to failures in satisfying budget constraints. Robust identification of the main product region poses another challenge for current agents.
    
    \item \textit{\textbf{Insufficient exploration.}} 
    We find a gap between autonomous web agents' internal evaluation and search behavior. They sometimes correctly recognize when products fail to meet requirements, such as exceeding budget or unavailable shipping to the user's location. Yet, they often explore only a single platform and terminate without searching other sources (Figure ~\ref{fig:insufficient_adaptation}). This indicates that autonomous web agents need to be more encouraged to conduct broader exploration when initial options fail.
\end{itemize}

\end{document}